# Charge ordering as the driving mechanism for superconductivity in rare-earth nickel oxides

Álvaro Adrián Carrasco Álvarez[1,2], Lucia Iglesias[2], Sébastien Petit[1], Wilfrid Prellier[1], Manuel Bibes[2] and Julien Varignon[1]

[1] CRISMAT, ENSICAEN, Normandie Université, UNICAEN, CNRS, 14000 Caen, FRANCE
[2] Unité Mixte de Physique, CNRS, Thales, Université Paris Saclay, 91767 Palaiseau, France

**Abstract**

**Superconductivity is one of the most intriguing properties of matter described by an attractive interaction that binds electrons into Cooper pairs. To date, the highest critical temperature at ambient conditions is achieved in copper oxides. While layered nickel oxides were long proposed to be analogous to cuprates, superconductivity was only demonstrated in 2019 albeit without clarifying the pairing mechanism. Here we use Density Functional Theory (DFT) to show that superconductivity in nickelates is driven by an electron-phonon coupling originating from a charge ordering. Due to an intrinsic electronic instability in half-doped compounds, $Ni^{1.5+}$ cations dismutate into more stable $Ni^+$ and $Ni^{2+}$ cations, which is accompanied by a bond disproportionation of $NiO_4$ complexes producing an insulating charge ordered state. Once doping suppresses the instability, the bond disproportionation vibration is sufficient to reproduce the key characteristic of nickelates observed experimentally, notably the dome of $T_c$ as a function of doping content. These phenomena are identified if relevant degrees of freedom as well as an exchange correlation functional that sufficiently amends self-interaction errors are involved in the simulations. Finally, despite the presence of correlation effects inherent to *3d* elements that favour the formation of local Ni spins, the mechanism behind the formation of Cooper pairs in nickelate superconductors appear similar to that of non-magnetic bismuth oxide superconductors.**

Superconductivity is a peculiar property of materials characterized by zero-electrical resistance to direct current and perfect diamagnetism, thereby offering numerous applications. This state is explained on the basis of an attractive interaction binding electrons into Cooper pairs [1]. The most famous superconductors are cuprates that set the record for the critical temperature $T_c$ (~138 K) at ambient conditions[2]. However, the identification of the pairing mechanism in these compounds has been hindered by the absence of analogues. In this regard, it was long proposed that nickel oxides could host superconductivity due to their proximity with copper in the periodic table[3,4]. These efforts recently crystallized in infinite layered phase $RNiO_2$ [5–8] and a reduced phase of the Ruddlesden-Popper series (RRP) $R_{n+1}Ni_nO_{2n+2}$ (n=5)[9], where R is a trivalent rare-earth, thereby offering an alternative playground to understand pairing mechanisms in oxide superconductors.

Superconducting nickelates adopt the chemically reduced Ruddlesden-Popper (RRP) structure $R_{n+1}Ni_nO_{2(n+1)}$, where the most studied case is the limit member n=∞ (called the infinitely layered phase $RNiO_2$) (**Figure 1.a**). These nickelates are based on $NiO_2$ planes (**Fig. 1.a**) stacked along the *c* axis and separated by layers of rare-earth R. The motif is then sandwiched with two $RO_2$ fluorite slabs, except when n=∞ for which only R cations are intercalated between the $NiO_2$ planes. This layered structure of $NiO_2$ planes is common to several oxide superconductors such as cuprates[2], ruthenates $Sr_2RuO_4$[10] or iridates $Sr_2IrO_4$[11,12].

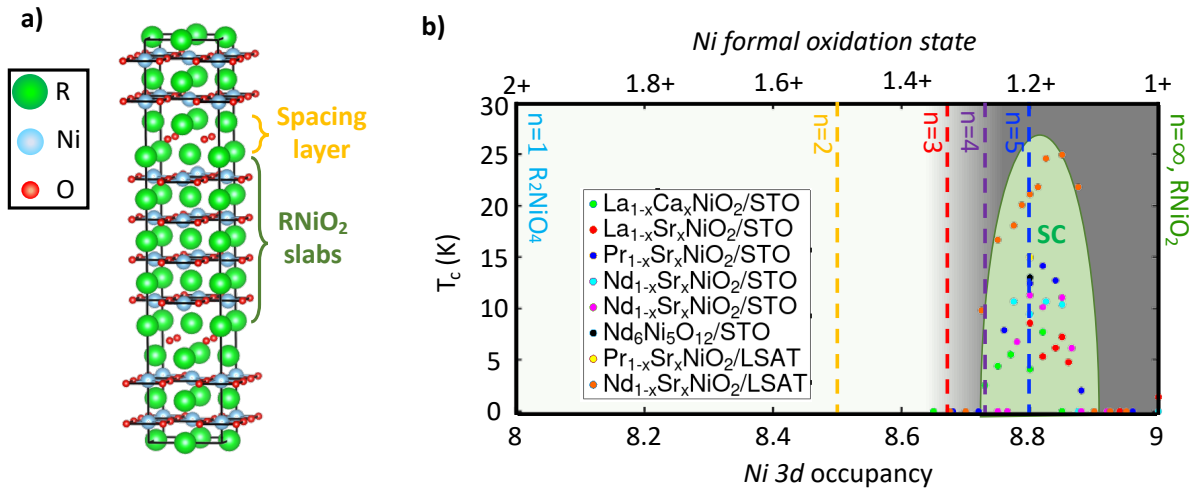

*Figure 1: Structural and electronic properties of nickelates. (a) Reduced Ruddlesden-Popper (RPP) phase exhibited by nickelates consisting of n $RNiO_2$ slabs intercalated between spacing layers. (b) Critical temperature $T_c$ (in K) of rare-earth nickelates versus the Ni 3d electron count extracted from experimental results available in literature. Experimental critical temperatures $T_c$ are taken from Refs.*[5–9,13].

Gathering the experimental data from literature regarding the $T_c$ versus Ni *3d* electron count (**Figure 1.b**), it becomes clear that superconductivity emerges in a narrow Ni valency region, centered at a $3d^{8.8}$ occupancy. This strongly suggests that the Ni valency is the determining factor in the transition towards the SC phase. Most studies in literature focus on how transiting from the n=∞ member to the SC region by hole doping the material[14–22]. However, these n=∞ nickelates are experimentally found to be bad metals[5], as also found by our DFT calculations (see **Appendix Figure 1**). This strategy is quite surprising and it is at odds with the usual procedure to reach the SC region in several oxides superconductors such as cuprates, bismuthates or antimonates where doping induces the transition from an insulating state to a metallic, superconducting phase.

This apparent contradiction is alleviated if one reads the phase diagram of **Figure 1.b** starting from the n=1 member. In that regard, studying the electronic properties of $La_2NiO_4$, we find it to be a robust Mott insulator at the DFT level with a large band gap of 2 eV (**see Appendix Figure 1**). Thus, reaching the superconducting phase starting from $La_2NiO_4$ requires to dope with electrons a Mott insulator. Therefore, one may ask if "one can *interpret the superconducting phase of from the point of view of electron doping Mott insulators instead of hole doping metals*". If this is true, then "*can we get insights on the superconducting mechanism in these nickel oxide superconductors?*". For instance, "*is there any intrinsic instability in the doped phase diagram of these nickelates?*".

In this article, we reveal on the basis of Density Functional Theory (DFT) simulations the existence of a charge ordered (CO) insulating state at half doping. The CO is a consequence of an unstable 1.5+ formal oxidation state (FOS) of Ni cations that prefers to adopt the 1+ and 2+ FOS in the ground state, either as a RRP n=2 or as a $La_{0.5}Sr_{0.5}NiO_2$ infinite layered structure. The CO phase is accompanied by a bond disproportionation producing compressed and extended $O_4$ complexes. Using this insulating state as a starting point, we then reveal that doping the material with electrons progressively suppress the instability toward disproportionation effects. Then the distortion is completely suppressed once we reach an effective Ni valency of 8.75, reminiscent of the valency required to reach SC experimentally. Thus, it appears that SC is at the vicinity of a charge ordered state, in close similarity of bismuth and antimony oxide SCs[23–26]. When the metallic state is reached, we then show that the disproportionation vibration is sufficient to reproduce the experimentally observed critical temperatures of $La_{1-x}Sr_xNiO_2$. Thus, superconductivity is dominated by an electron-phonon coupling in these nickelates, despite the presence of correlation effects inherent to transition metal cations. All these results are obtained by involving all relevant degrees of freedom such as structural lowering events and local spin formation as well as an appropriate exchange-correlation functional. Most notably, the usual non spin-polarized (NM) is shown to fail at capturing the bond disproportionation effects and to produce energy solutions at least 150 meV (*i.e.* 1800 K) greater than solutions with an appropriate paramagnetic representation of Ni spin interactions.

**METHODOLOGY**

Prior to discussing the physical properties of electron doped nickelates, it is relevant to discuss several critical elements of the *first-principles* simulations. Literature abounds of DFT based studies of the superconducting nickelates using a non-spin polarized (NM) solution, with symmetry conserving undistorted cells and/or in combination of a poor description of exchange-correlation (*xc*) phenomena, such as involving Local Density Approximation or Generalized Gradient Approximation (GGA) density functionals. These critical points notably led to rule out an electron-phonon coupling in $RNiO_2$ from a combination of NM/GGA calculations in Ref.[15] and DFT/LDA in Ref.[27], the latter highlighting the importance of treating correlation effects at the prohibitive Green's function (GW) level. The reason behind these shortcuts comes from limited possibilities for computing the full electron-phonon coupling from *first-principles* simulations: (i) theoretical tools can only be performed at a non-spin-polarized level so far and (ii) the calculation is far too demanding in terms of computational resources for including an appropriate *xc* functional and allowing symmetry lowering events (larger cell with less symmetry). Although a fantastic tool for computing electron-phonon response in non-magnetic systems with weakly correlated features, it is however not suited for studying complex oxide systems involving transition metal elements. Notably, Ref.[28] demonstrated that LDA/GGA *xc* functionals as well as the NM approximation fall outside the appropriate possibilities for describing properties of transition metal oxide materials. Finally, as we discuss below, a practical implementation for estimating superconducting quantities with band structure techniques goes through the estimation of band splitting associated with given lattice distortions/vibrations. Within this procedure, superconducting quantities are related to quantification of band splitting and hence are extremely sensitive to the choice of the DFT *xc* functional.

***Superconducting properties***: We have calculated the reduced electron-phonon matrix element (REPME) associated with a distortion B by freezing its atomic displacement in the relaxed ground state. Using the gap amplitude $\Delta E_k$ appearing in the band structure at the k coordinate in the first Brillouin zone due to the frozen phonon displacement, we compute the REPME by using the following formula $D = \frac{\Delta E_k}{2u}$, where $u$ is the displacement of one O atom for the condensed phonon mode – if it does involve only O motions. This standard procedure was successful in $BaBiO_3$, $SrBiO_3$ or $MgB_2$ compounds[25,29,30]. In order to calculate the electron-phonon coupling $\lambda$, we use the following formula $\lambda = N(E_F) \frac{\hbar^2}{2M\omega_B^2} D^2$, where $N(E_F)$ is the density of states at the Fermi level per spin channel and per formula unit, $M$ is the mass of the moving ion in the phonon mode and $\omega_B$ is the frequency of the B phonon mode. To obtain the critical temperature $T_c$, we use the Mc Millian-Allen equation[31] $T_c = \frac{\omega_{log}}{1.12} \exp\left(\frac{-1.04(\lambda+1)}{\lambda - \mu^*(1+0.62\lambda)}\right)$, where $\omega_{log}$ is the logarithmic averaged phonon frequency (expressed in K) and $\mu^*$ is the screened Coulomb potential with conventional values ranging from 0.1 to 0.15.

***The choice of the DFT xc functional***: LDA and GGA cannot sufficiently amend self-interaction errors inherent to practiced DFT, particularly for *3d* transition metal elements. Therefore, it usually yields metals while most *3d* transition metal compounds are insulators experimentally[28,32]. Since the practical implementation for estimating superconducting quantities relies on the estimation of band splitting in band structures in the present study, it should be modeled with care. Involving more evolved DFT functionals such as meta GGA or Hybrid may help to describe accurately the electronic features and superconducting properties. This was already highlighted in Ref.[25,29]. In this study, most simulations are performed with the Strongly Constrained and Appropriately Normalized (SCAN) meta-GGA functional[33]. This functional is able to predict the correct trends in lattice distortions and metal-insulator transitions in bulk $ABO_3$ perovskite oxides and trends in doping effects in nickelates and bismuthates[25,32,34]. In addition, SCAN does not require any external parameter as in DFT+U and hence is able to adapt to different oxidation states of ions as a result of doping or appearance of charge orderings. However, this functional may underestimate the band gap of these strongly correlated nickel oxide with a low nominal valence – the lower the formal oxidation state, the larger the self-interaction errors[35]—, as in highly uncorrelated semiconductors. We have thus benchmarked our DFT-SCAN results by performing one shot simulations on smaller nickelate systems with a hybrid HSE06[36] functional to evaluate the improvement on band gap properties, in the spirit of Ref.[29]. Although the hybrid functionals can sometimes over localize electrons and open band gaps in materials that are metals experimentally such as in the doped cuprates[37,38], the HSE06 functional has been already used in the infinite layered nickelates and shown to preserve the metallic character of the limiting members $LaNiO_2$ and $PrNiO_2$[35]. The standard parametrization of the HSE06 functional was used, namely with 25% of exact Hartree-Fock exchange and a range separation of 0.2 of Å$^{-1}$. We however emphasize that all calculations have been performed with the SCAN functional unless stated.

***Local spin formation may not be overlooked***: Within the NM representation, all transition metal elements are forced to have no magnetization at all, *i.e.* $n_\uparrow = n_\downarrow$ where n is the number of electrons with spin ↑ or ↓. This is quite surprising for elements such as Ni exhibiting a *3d⁹* configuration with a spin-unpaired electron in the infinite layered $RNiO_2$ phase. Alternatively, one may use a supercell with a true paramagnetic order by randomizing spins at the restriction that the sum over all spins must be zero. Using this representation, we indeed identify that the energy difference between a fully relaxed NM and PM in our DFT simulations as a function of doping content in $La_{1-x}Sr_xNiO_2$ phase diagram is massive, with at least 150 meV/f.u of energy gain by developing a spin-polarized solution (**Appendix Figure 2**). It corresponds to a NM phase that would be stable above at least at T=1800 K. This value has to be compared with the 15-25 K targeted for the critical temperature $T_c$ of superconducting nickelates. As we will show later in the manuscript, neglecting the Hund's rule coupling in the DFT simulation has several dramatic consequences as it prevents structural lowering events strongly coupled to electronic features and enables a totally different band structure problem. Thus, the NM approximation is not an appropriate starting point for understanding the superconducting mechanism. Even though electron-electron interactions are treated with the highest level of description such as with DFT with hybrid functionals or by introducing many body Green's function for instance, the results are dependent on the input basis and their application on a non-spin polarized input will only produce an improvement of the band dispersion of a non-spin polarized system.

To circumvent potential spin-phonon couplings, we used a paramagnetic representation for Ni spins following the strategy presented in Ref.[28]. The ATAT package[39] and the Special Quasi random Structures (SQS)[40] are used for identifying the spin arrangement maximizing the disorder characteristic of a random spin configuration within a given supercell size, albeit spins are only treated at the colinear level.

***Crystallographic cells, structural relaxations and analysis***: The imposed starting cells correspond to high symmetry $P_4/mmm$ and the $I_4/mmm$ tetragonal unit cells for the $RNiO_2$ infinitely layered phases and the reduced Ruddlesden-Popper phase, respectively. Solid solutions of the infinite layered phase corresponds to a $(2\sqrt{2},2\sqrt{2},4)$ supercell with respect to the primitive $P_4/mmm$ cell (32 formula unit). The RRP n=2 phase is modelled by a (2,2,1) supercell, corresponding to a supercell with 176 ions. The choice of these cell is guided by their ability to accommodate lattice distortions as well as the PM solution. We then proceed to the full structural relaxation (cell parameters and shape as well as atomic positions) until the forces acting on each atoms are lower than 0.05 eV/Å. Amplitude of the distortions of the relaxed ground states are then extracted using symmetry mode analysis taking as reference the undistorted $P_4/mmm$ and $I_4/mmm$ cells for infinitely layered and RRP phases, respectively. This is performed with the ISODISTORT tool from the ISOTROPY applications[41,42].

***Doping effects in superlattices and disordered solid solutions***: Doping effects are performed by substituting $La^{3+}$ cations by $Sr^{2+}$ divalent cations. In the Disordered Solid Solution (DSS), the disorder characteristic of an alloy is extracted using the Special Quasirandom Structure (SQS) technique developed by Zunger *et al*[40], allowing to extract the cation arrangement maximizing the disorder and the occurrence of all possible local motif in a given supercell size. The DSS cell size is limited to 32 f.u. in order to have a feasible calculation with PM and/or cation disorder inducing a *P1* symmetry.

*Seeking for electronic instabilities:* The SCAN functional is a local functional of the density matrix unable to make distinction between occupied and unoccupied states. It is therefore unable to identify electronic instabilities in high symmetry cells with degenerate partners as performed in Ref.[28] where one has to impose integer occupancy of a specific degenerate partner such (1,0) instead of (0.5,0.5) for two degenerate orbitals. Instead we used the strategy proposed in Ref.[43]: we plot the potential energy surface associated with a lattice distortion and seek to see the shape of the potential energy surface. A shifted single potential whose minimum is located at non zero amplitude of the mode then indicates the presence of an electronic instability as those observed for the Jahn-Teller or bond disproportionation distortions in Ref.[43].

Checks of the existence of electronic instabilities has nevertheless been performed on simpler systems with a ferromagnetic order of Ni spins by performing two sets of calculations on a high symmetry undistorted cell using hybrid DFT HSE06 calculations: (i) one calculation where one enforces similar occupancies of degenerate partners and a (ii) a calculation where one imposes specific orbital occupancies and switching off the symmetry on the wavefunction. The difference of energy between calculations (i) and (ii) then indicates the presence or absence of electronic instabilities at 0K.

*Potential energy surface and phonon frequencies*: The potential energy surfaces associated with a lattice distortion labelled B are computed taking as a reference the undistorted *P4/mmm* and *I4/mmm* for the infinitely layered and the reduced Ruddlesden-Popper phase, respectively. Then a finite distortion amplitude $Q_B$ is frozen in the material and the energy is computed for different amplitudes. Regarding the phonon frequencies, a full phonon calculation is not affordable since the DSS formula unit supercell contains more than 100 atoms without any symmetry. Instead, we simply compute the potential energy surface of the desired distortion in the relaxed ground state structure by freezing different mode amplitude as proposed in Ref.[25]. Due to the presence of an electronic instability that moves the single well minimum to non-zero amplitude of $Q_B$, the evolution of the energy ΔE is given by the following expression:

$$\Delta E = E_0 + \alpha(Q_B - Q_0)^2 + \beta(Q_B - Q_0)^4 \quad \textbf{(eq.1)}$$

where $\alpha$ and $\beta$ are coefficients and $Q_0$ signals the force acting on the electrons even in the absence of $Q_B$ (*i.e.* the electronic instability). It follows that

$$\Delta E = E_0 + \alpha Q_B^2 + \alpha Q_0^2 - 2\alpha Q_B Q_0 + \beta Q_B^4 + \beta Q_0^4 - 4\beta Q_B^3 Q_0 - 4\beta Q_B Q_0^3 + 6\beta Q_B^2 Q_0^2 \quad \textbf{(eq.2)}$$

$$\Delta E = E_0 + (\alpha Q_0^2 + \beta Q_0^4) + (-2\alpha Q_0 - 4\beta Q_0^3) Q_B + (\alpha + 6\beta Q_0^2) Q_B^2 + (-4\beta Q_0) Q_B^3 + \beta Q_B^4 \quad \textbf{(eq.3)}$$

$$\Delta E = E_0' + b\, Q_B + c Q_B^2 + d Q_B^3 + e\, Q_B^4 \quad \textbf{(eq.4)}$$

with $E_0' = E_0 + (\alpha Q_0^2 + \beta Q_0^4)$. From **eq.3**, we recover the linear and trilinear term in $Q_B$ signaling the contribution of the electronic instability. Using a fit of the potential energy surfaces as a function of $Q_B$ with a polynomial expression up to the 4$^{th}$ order, we can extract all coefficients of the **eq.4** and map them to get the mode frequency through the relation $\omega = \sqrt{\frac{2\alpha}{M}}$ -- *i.e* $E_{harm} = \frac{1}{2} M \omega^2 Q_B^2$.

***Other technical details***: DFT simulations are performed with the Vienna Ab initio Simulation Package (VASP)[44,45]. Projector Augmented Wave pseudo potentials[46] (PAW) are used taking the $3d^8 4s^2$, $5p^6 5d^1 6s^2$ and $4s^2 4p^6 5s^2$ electrons as valence electrons for Ni, La and Sr atoms, respectively. The energy cut-off is set to 650 eV and is accompanied by a 3x3x2 Gamma centered k-mesh for the 32 f.u. supercells. The k-mesh is set to 3x3x1 for the RRP n=2 cell (176 ions/supercell). Wannier functions are built using the wannier90 package[47–50]. In order to have accurate Density of states at the Fermi level, we have employed the wannier90 package[47–50] on top of our electronic structure calculation. Using the Bloch states four our relaxed 32 f.u supercells, we extract the density of states on very dense kmesh of 64x64x64 points.

**RESULTS**

***Charge disproportionation and insulating state at half-doping:*** The next member in the doped nickelate phase diagram is for n=2 member, which correspond to a half-doped material. We thus inspect the structural and electronic properties of half-doped nickelates $La_{0.5}Sr_{0.5}NiO_2$ either as a superlattice (SL) with perfect layering between La and Sr layers along the c axis or a purely disordered solid solution (DSS) and also in the "self-doped" Reduced Ruddlesden-Popper (RRP) phase $La_3Ni_2O_6$. All systems possess a $Ni^{1.5+}$ - $3d^{8.5}$ electronic configuration. After structural relaxation using a paramagnetic solution, we identify that the ground state of these materials is characterized, among all emergent lattice distortions, by a bond disproportionation mode $B_{oc}$ – *e.g.* breathing of oxygen complexes — splitting the Ni sites into cations sitting in compressed ($Ni_S$) and extended ($Ni_L$) oxygen complexes and forming a layered checkerboard structure (**Figure 2.a**). Using a symmetry mode analysis with respect to the primitive undistorted cell, we find sizable amplitudes of 0.0719, 0.0603 and 0.089 Å/NiO$_2$ motif in the SL, DSS and RRP phases, respectively. The structural relaxations are associated with large energy gains with respect to the undistorted phases of -87, -53 and -32 meV/$NiO_2$ motif for the SL, DSS and RRP phases, respectively.

The effect of the $B_{oc}$ mode is to produce a clear asymmetry of the electronic structure associated with the two different Ni cations in the cell as inferred by the projected density of states (pdos) on $Ni_L$ and $Ni_S$ cations (**Figure 2.b**) for the half-doped SL. The Fermi level physics is dominated by $Ni_L$ *d* states while $Ni_S$ *d* states are pushed above $E_F$ and separated from the $Ni_L$ states. Furthermore, the O *p* orbitals do not bring sizable contribution to the DOS in this energy range and the top of these states are in fact located 1 eV below the Fermi level. By looking at the partial charge density map originating from bands just at the Fermi level $E_F$, a charge ordering between $Ni_L$ and $Ni_S$ cations (**Figure 2.c**) is identified with $Ni_L$ likely holding more electrons than $Ni_S$ cations. The partial charge density maps also highlights the strong Mott character of these nickelates with negligeable spread of the electronic structure on O atoms. The Mott character is in agreement with previous results on undoped materials[51–55]. In addition, the existence of a charge ordering is confirmed by our computed magnetic moments revealing that $Ni_L$ sites bear a magnetic moment of 0.99±0.05 $\mu_B$ depending on the nature of the half-doped nickelate (*i.e.* SL, DSS or RRP) and $Ni_S$ sites have a moment of 0.26±0.18 $\mu_B$. The slight deviation from 0 $\mu_B$ for $Ni_S$ sites comes from cation and/or spin disorders yielding a different local potential experienced by each Ni cations – see **Appendix Figure 3**. Such a deviation was already observed in the DFT-PM state of disproportionating materials in Ref.[28] The clear magnetic moments asymmetry suggests that $Ni_L$ cations are in a $Ni^{1+}$ – *$3d^9$* configuration while $Ni_S$ cations are in a $Ni^{2+}$ – *$3d^8$* low spin state.

Aiming at understanding how electrons localize in the ground state, we have built the Wannier functions (WFs) associated with the ground state of the 50% doped SL following the strategy presented in Ref.[56] After minimizing the spread of the WFs, we infer that (i) all O *p* states are doubly occupied; (ii) for $Ni_L$ cations all *d* orbitals are doubly occupied except the $d_{x^2-y^2}$ orbital that holds a single electron and (iii) for $Ni_S$ cations all *d* orbitals are doubly occupied except the $d_{x2-y2}$ orbital that is empty (see **Appendix Figure 4**). It confirms that the ground state of half-doped nickelates is associated with a charge order between $Ni_L^+$ – $3d^9$ and $Ni_S^{2+}$ – $3d^8$ low spin cations. We conclude here that nickelates with $Ni^{1.5+}$ cations are prone to exhibit a bond disproportionation $B_{oc}$ mode producing a charge ordering between $Ni^+$ – $3d^9$ and $Ni^{2+}$ – $3d^8$ cations.

As observed in the $ABO_3$ perovskites such as $RNiO_3$ (R=Lu-Pr,Y), $ABiO_3$ (A=Sr,Ba) or $BaSbO_3$ compounds[26,28,32,57,58], the effect of the $B_{oc}$ mode is to open a band gap in the material. Our SCAN calculation predicts a weakly metallic state for the DSS, SL and RPP with the PM order. However, this is inherent to DFT functionals that are local or semi-local operators of the non-interacting density matrix (*e.g.* LDA, GGA) that underestimate the band gap of compounds, including those of highly uncorrelated semi-conductors. Using a more accurate but prohibitive HSE06 hybrid functional on short range spin ordered half doped heterostructures yields an insulating state (see **Appendix Figure 5**).

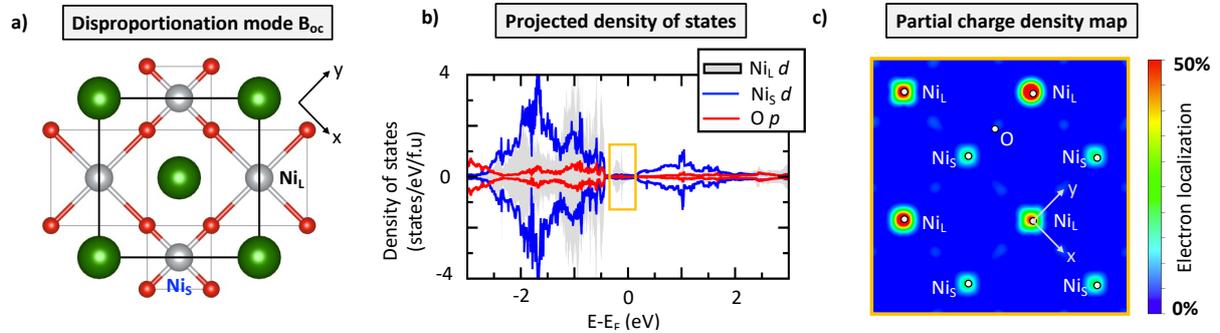

***Figure 2 : Key properties of half-doped 2D nickelates. a)*** *Bond disproportionation $B_{oc}$ mode splitting Ni sites into $Ni_L$ and $Ni_S$ sitting in an extended and compressed oxygen complex, respectively.* ***b)*** *Projected density of states on $Ni_L$ (filled grey) and $Ni_S$ (blue line) d states and O p states (red line) for the 50% doped superlattice.* ***c)*** *Partial charge density map associated with bands just below the Fermi level indicated by the orange square in panel **c** in the half-doped superlattice.*

In order to reveal the origin of the disproportionation effects, we examine the potential energy surface (PES) associated with the disproportionation mode $B_{oc}$ in the SL, DSS and RRP. To that end, we compute the total energy by freezing some amplitude of the $B_{oc}$ mode starting from a high symmetry undistorted cell (**Figures 3**). In all three forms of half-doped nickelates, we identify a single well potential whose minimum is shifted to non-zero amplitude of the $B_{oc}$ mode. This indicates the existence of a force associated with an electronic instability acting to remove the electronic degeneracy of the $3d^{8.5}$ electronic configuration[43]. Furthermore, the instability is independent of the form of the structure (*i.e.* SL *vs.* RRP) as well as of the order/disorder of A site cations (*i.e.* SL *vs.* DSS). This suggests that the formal occupancy of Ni *3d* orbitals is the determining factor rather than the form of the nickelate. The existence of an electronic instability associated with the unstable 1.5+ formal oxidation state (FOS) of Ni cations is confirmed by an hybrid HSE06 calculation on an undistorted cell for which a $d^8/d^9$ charge ordering is more stable by 11.5 meV/f.u. than a $d^{8.5}$ occupancy of all Ni cations (see **Appendix Figure 6**) and already open a finite band gap in the system. We conclude here that the 1.5+ FOS of Ni cations is intrinsically unstable and is willing to disproportionate to more stable 1+ and 2+ FOS in the ground state, thereby producing a charge ordering and a bond disproportionation mode $B_{oc}$.

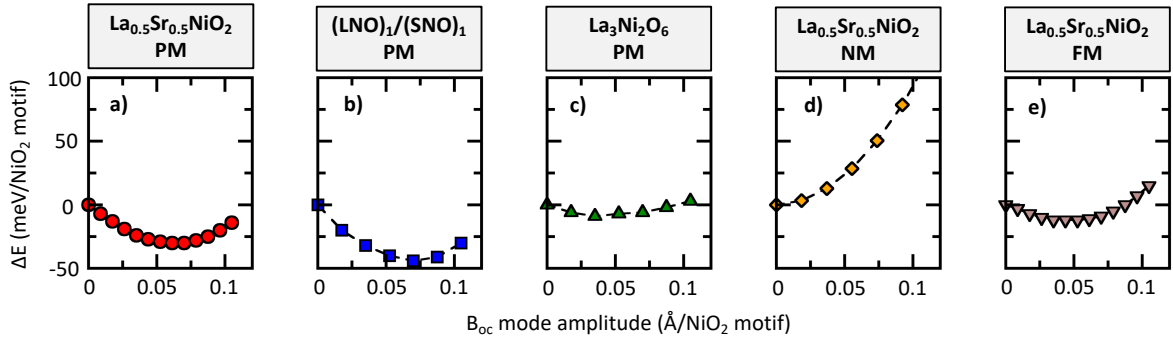

*Figure 3: Origin of disproportionation effects. a-c)* Total energy gain (in meV/NiO$_2$ motif) as a function of the disproportionation B$_{oc}$ mode amplitude (in Å/NiO$_2$ motif) for the half-doped nickelates using a solid solution (**a**), a superlattice (**b**) and the n=2 reduced Ruddlesden-Popper phases (**c**) with paramagnetic order. *d-e)* Total energy gain (in meV/NiO$_2$ motif) as a function of the disproportionation B$_{oc}$ mode amplitude (in Å/NiO$_2$ motif) for the half-doped nickelates using a NM (**d**) and FM (**e**) order. The point at 0 amplitude is set as the reference energy and corresponds to an undistorted cell.

The existence of a disproportionated insulating state at half doping is confirmed experimentally in La$_3$Ni$_2$O$_6$ (RRP n=2)[59,60]. This was also predicted theoretically in Ref.[61]. Nevertheless, the existence of a half doped infinite layered phase is yet to be realized experimentally, although the phase stability may be questionable. However, should such a phase exist, it would be characterized by robust disproportionation effects of Ni$^{1.5+}$ cations.

***Local spin formation is a key degree of freedom:*** Allowing spin degree of freedom and satisfying the basic Hund's rule for Ni cations is crucial since a non-spin polarized (NM) DFT calculation yields a single well potential for B$_{oc}$ whose minimum is centered at zero amplitude (**Figure 3.d**). In contrast, all additional calculations performed with various long-range spin orders show the stabilization of the mode (see **Figure 3.e** for the FM case). This is in agreement with the existence of the breathing mode B$_{oc}$ in the perovskite RNiO$_3$ phase that requires the Hund's rule to be respected[32] (*i.e.* formation of local spins). Nevertheless, disproportionation effects are unrelated to correlation effects as it is extensively discussed in Refs.[25,28,32,58]. Thus, along with predicting solution at 1800 K/f.u above a true PM solution, the NM approximation cannot capture the relevant structural lowering event associated with intrinsic electronic instabilities. It is thus irrelevant for studying the physics of nickelate superconductors.

***Doping is a lever to suppress the charge ordered insulating state:*** Having established the existence of disproportionation effects in half-doped nickelates and of an associated insulating phase, we now use this result as a starting point for studying the role of doping – *i.e.* hole doping starting from pristine LaNiO$_2$ material or electron doping starting from the half-doped La$_{0.5}$Sr$_{0.5}$NiO$_2$. The insulating starting point being similar between all forms of superconducting nickelates and a common point with the rest of the SC oxides. We now restrict the study to the DSS for convenience. To that end, we decrease the x value from 0.5 in La$_{1-x}$Sr$_x$NiO$_2$. We report in **Figure 4.a** the trend in bond disproportionation amplitude and magnetic moment asymmetry between Ni$_L$ and Ni$_S$ sites as a function of the nickel 3d electron count. We observe that starting from Ni$^{1.5+}$ cations (x=0.5, Ni $3d^{8.5}$), the bond disproportionation amplitude extracted from a symmetry mode analysis decreases until x=0.25 (Ni $3d^{8.75}$). For x<0.25, the mode is no longer stable. The same conclusions are raised using the magnetic moments difference displayed in **Figure 2.b** as well as with a long-range AFM order (see **Appendix Figure 7**). The extinction of the disproportionation effects at x=0.25 is reminiscent of the value at which superconductivity emerges experimentally in these compounds (x~0.2 in DSS or n=5 in RRP phases, corresponding to Ni$^{1.2+}$ - $3d^{8.8}$ cations)[25]. We emphasize that although using a simpler AFME order compatible with the appearance of B$_{oc}$ modes in oxides[43], the doped case at 37.5% ($3d^{8.625}$) is found insulating with HSE06 DFT functional (see **Appendix Figure 8**). We conclude here that doping acts to suppress the disproportionation effects – and the insulating phase — present in the half-doped nickelates.

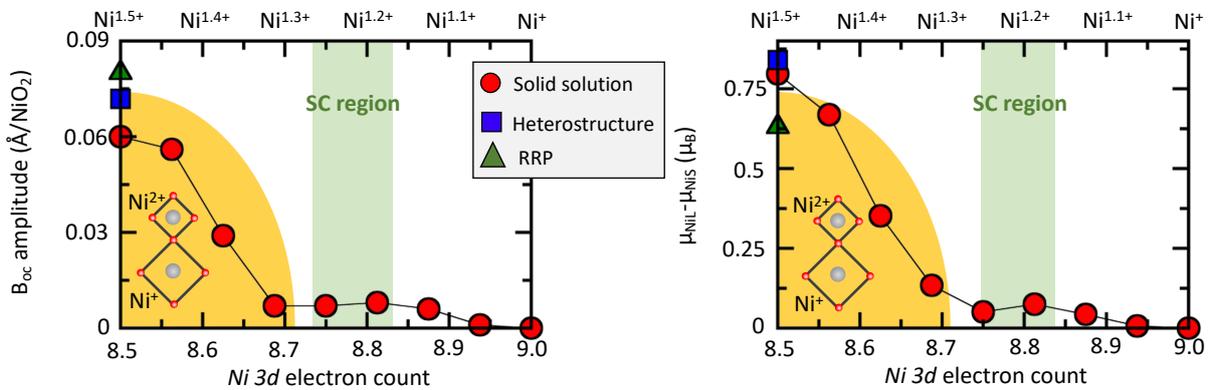

*Figure 4 : Trend in disproportionation effect with doping content. Amplitude (in Å/NiO$_2$ motif) associated with the B$_{oc}$ bond disproportionation mode (**a**) and asymmetry of Ni$_L$ and Ni$_S$ magnetic moments (**b**) as a function of Ni 3d electron count (lower scale) or Ni formal oxidation state (upper scale) for the fully relaxed ground state superlattices (SL, filled blue squares), disordered solid solutions (DSS, red filled circles) and reduced Ruddlesden-Popper (RRP, filled green diamonds).*

Previous experimental works on RRP phase with n=3 revealed the existence of a charge ordered state (Ni$^{1.33+}$ - $3d^{8.66}$)[62–64], as well as an insulating character below 105 K[63,65], in agreement with our simulations. These findings are also supported by various theoretical works on the very same compounds showing that the ground state is formed of Ni$^+$ and Ni$^{2+}$ cations[66], in line with our results for the doped solid solution. Finally, an earlier theoretical work also revealed the propensity of the hole doped NdNiO$_2$ to develop bond/charge ordered states[67]. We finally emphasize that oxygen rich NdNiO$_{2+\delta}$ samples are also prone to exhibit charge orderings[68,69], highlighting the propensity of these doped infinite layered nickelates to develop charge disproportionation effects.

***Vicinity of a charge ordered state is a prerequisite to SC:*** We report in **Figure 5.a** the potential energy surface of the B$_{oc}$ mode starting from a high symmetry cell for different doping contents for the DSS. Starting from the ideal doping content x=0.5 (Ni $3d^{8.5}$), we see a shifted single well down to x=0.25 (Ni $3d^{8.75}$), albeit the minimum is progressively displaced to lower mode amplitudes. Further decreasing *x* transforms the shifted single well to a single well potential whose minimum is located at zero amplitude of the mode, indicating that the electronic instability associated with disproportionation effects is suppressed. At this stage, no more charge orderings – and hence semi-conducting states – are expected. From x=0.25 (Ni $3d^{8.75}$) to x=0 (Ni $3d^9$), the curvature of the total energy as a function of the amplitude of displacement Q$_{Boc}$ around the origin becomes steeper. Aiming at extracting the phonon frequency as a function of the doping content, we have fitted the PES of **Figure 5.a** as a function of Q$_{Boc}$ with a polynomial expression of the form $\Delta E = \alpha(Q_{Boc} - Q_0)^2 + \beta(Q_{Boc} - Q_0)^4$, where $Q_0$ corresponds to the shift of the energy minimum induced by the force associated with electronic instability (see methods for further details) and $\alpha$ and $\beta$ are coefficients. Recalling that the harmonic contribution to the energy is $E_{harm} = \frac{1}{2}M\omega^2 Q_{Boc}^2$ where M is the mass of the moving atoms, we directly get that $\omega^2 = \frac{2\alpha}{M}$.

From the extracted squared B$_{oc}$ phonon frequency as a function of doping content reported in **Figure 5.b**, we see that $\omega_{Boc}^2$ strongly hardens from x=0.25 to x=0. Above x=0.25, the B$_{oc}$ mode frequency without the electronic instability would increase. It follows that the B$_{oc}$ mode reaches its lowest frequency $\omega_{Boc}$ at the vicinity of a stable charge ordered phase boundary, thereby maximizing the probability of its vibration in the material. In turn, it possibly favors the punctual formation of coupled electrons and holes on the lattice, *i.e.* the Cooper pairs. Although $\omega_{Boc}^2$ is rather low above x=0.25, electrons and holes are localized in the lattice due to the finite stabilization of the B$_{oc}$ mode induced by the electronic instability as discussed above in the manuscript, and *de facto* superconductivity cannot exist. A totally different picture is achieved with the NM approximation that provides a monotonous strengthening of $\omega_{Boc}$ with increasing the doping content (**see Appendix Figure 9**), in addition to the absence of tendency toward charge disproportionation/ordering. Finally, we observe a very slight shift of $\omega_{Boc}^2$ to higher energies when considering the ground state structure (**Figure 5.b**). It signals the absence of significant couplings between the B$_{oc}$ mode and other lattice distortions, in contrast to the usual behavior observed in $RNiO_3$ and $ABiO_3$ compounds[25,28,70]. We conclude here that the vicinity of a charge ordered phase is a prerequisite for superconductivity in nickelates.

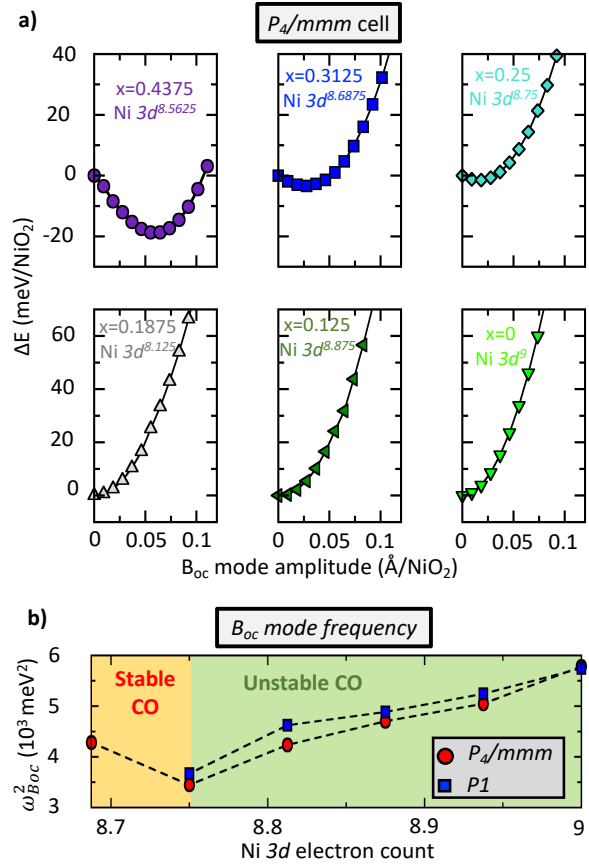

**Figure 5: Phonon frequency of the breathing mode $B_{oc}$ upon doping. a)** Energy gain $\Delta E$ (in meV/NiO$_2$) associated with the breathing mode amplitude (in Å/NiO$_2$) starting from an undistorted $P_4/mmm$ cell for the disordered solution $La_{1-x}Sr_xNiO_2$ at different doping contents using the PM order. The reference energy is set to the value at zero $B_{oc}$ mode amplitude. **b)** Extracted squared $B_{oc}$ phonon frequency (in meV$^2$) as a function of the Ni 3d electron count using the PM order within the undistorted $P_4/mmm$ (red filled squares) and real ground state structure (blue filled circles).

**Nickelates are associated with a single band system**: We report in **Figure 6.a** the unfolded band structure of the DSS in the doped region for $La_{0.8125}Sr_{0.1875}NiO_2$ with the PM order. We observe a single band system associated with the PM state in the SC region, described by a parabola centered at the A point -- (½, ½, ½) point of the Brillouin zone. This type of band structure is in sharp agreement with more complex Dynamical Mean Field Theory calculation[71] or non colinear PM DFT of Ref.[72]. The single band system agrees with the experimental data analysis performed by Talantsev[73] that through different adjustments determined that the nickelates superconductivity should be described by a single band. Again, this result contradicts the band dispersion obtained with a NM approximation which produces a two bands system (**Appendix Figure 10**) due to the absence of Hund's rule.

From the band structure projected on different atom and orbital contributions reported in **Figure 6.b**, the part of the parabola crossing the Fermi level along the Z-R-A path is dominated by Ni $d_{xz}/d_{yz}$ orbitals, with smaller contributions from La states. This is in sharp agreement with spin-polarized DFT simulations performed by Lane *et al*[74] or Zhang *et al*[54,75] that identify a parabola centered at the A point in undoped $NdNiO_2$ compounds with $d_{xz}/d_{yz}$ orbital character. The $d_{x2-y2}$ contributions that we observe do correspond to the impurity states introduced by the doping, and these orbitals are in fact located 1 eV below and 2 eV above $E_F$ in the undoped samples.

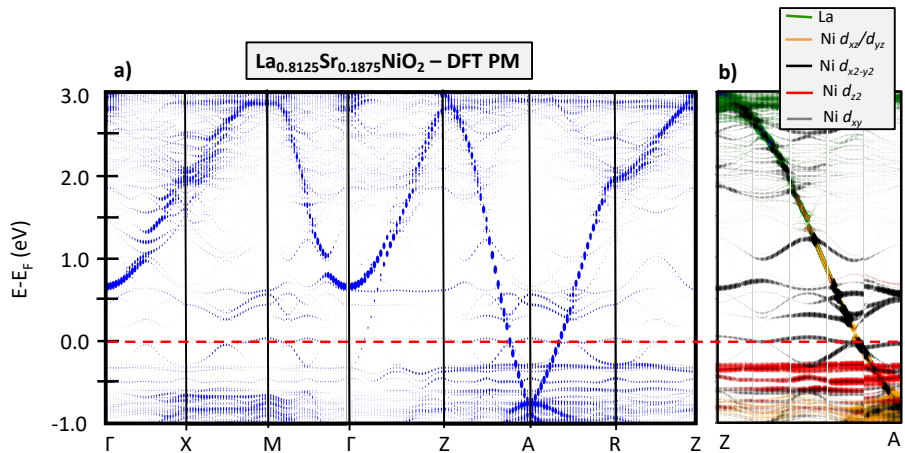

*Figure 6: Band dispersion associated of $La_{0.8125}Sr_{0.1875}NiO_2$. a) Unfolded band structure to the primitive high symmetry P4/mmm cell of $La_{0.8125}Sr_{0.1875}NiO_2$ with the PM order. Coordinates of the high symmetry points are the following: Γ(0,0,0), X(1/2,0,0), M(1/2,1/2,0), Z(0,0,1/2), R(1/2,0,1/2) and A (1/2,1/2,1/2). b) Bands dispersion along the Z-A path projected on all Ni $d_{xy}$ (grey), $d_{xz}+d_{yz}$ (orange), $d_{x2-y2}$ (black) and $d_{z2}$ (red) states and total contribution from the A cation (green) states.*

We report in **Figures 7.a-i** the unfolded band structures associated with the DSS with a PM solution as a function of the doping content *x* along the Z-A path –*i.e.* bands dispersing along the (0,0,½) – (½,½,½) and compatible with the (½,½,0) $B_{oc}$ mode symmetry. At x=0.5 ($3d^{8.5}$), an energy gap does exist in the band structure half-way of the distance to the Z and A points. This gap opening located 2 eV above the Fermi level is due to the presence of the bond disproportionation mode $B_{oc}$, a mode appearing at the (½,½,0) point of the Brillouin zone of the primitive cell – *i.e.* producing a cell doubling along the diagonal of the 2D planes. We emphasize that, the HSE06 that gives a better description of gap related properties, results in a clear gap opening at the Fermi level. When slightly electron doping the x=0.5 case (x=0.4375 and x=0.375), the bands are mostly rigid in energy but one notices that the gap appearing halfway the Z-A path decreases in amplitude until it totally disappears at x=0.25. This is in line with the disappearance of the $B_{oc}$ mode revealed from the symmetry mode analysis. From x=0.25 ($3d^{8.75}$) to x=0 ($3d^9$), one is left with a single parabola dispersing over roughly 3.6 eV and for which its intersection with the Fermi level is shifted when varying the doping content. It is worth noticing that a secondary gap located at 0.5 eV above $E_F$ and at quarter distance from the A point also emerge on some band structures (**Figure7.c** and **7.b**). However, it is likely a resulting effect of the spin disorder introduced in the material and only specific AFM patterns can introduce such a gap in the material (see **Appendix Figure 11**). We conclude here that bond disproportionation produces gaps in the band dispersion, a fact going against a fully dispersive band structure expected for a free electron gas and meaning that there is an interaction experienced by electrons from x=0.5 to 0.25.

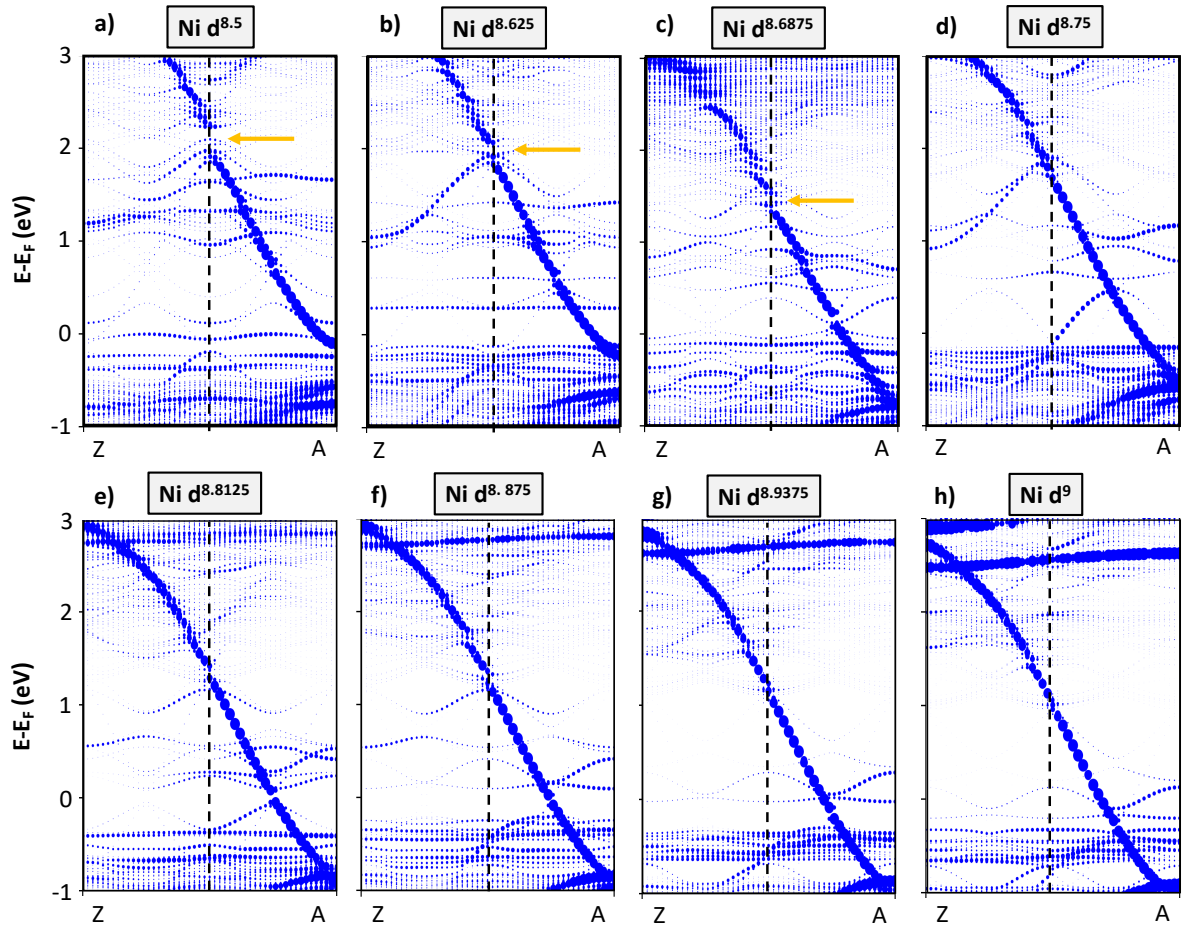

*Figure 7: Trends in band structures of nickelates as a function of the doping content.* Unfolded band structure of $La_{1-x}Sr_xNiO_2$ PM ground states to the Z-A path of the $P_4/mmm$ high symmetry Brillouin zone for different Ni 3d level occupancy and either with a disordered solid solution (DSS). Orange arrows indicate the gap opening coming from the $B_{oc}$ distortion. Size of dots are related to the contribution of folded bands within the primitive cell, and tiny points are mostly noise originating from the periodic replica of cells when distortions, cation and spin disorders are introduced in the supercell.

At this stage, one may sum up the different observations raised by the DFT simulations. A strong electron-phonon coupling exists in the half-doped regime and originates from an unstable 1.5+ FOS of Ni cations that prefers to transform into more stable 1+ and 2+ FOS in the ground state. It results in electron localization with the appearance of Ni$^+$ (3d$^9$)/Ni$^{2+}$ (3d$^8$ low spin) charge ordering that is accompanied by a Ni-O bond ordering (BO) through a breathing of oxygen complexes B$_{oc}$ distortion. It ultimately results in an insulating phase. Electron-doping the half-doped case alters the electron-phonon coupling albeit preventing the stabilization of a CO and BO until x=0.25 in La$_{1-x}$Sr$_x$NiO$_2$ compounds. For x<0.25, the electron-phonon coupling associated with the CO and BO is sufficiently weakened so that the no more electron localization is stabilized nor Ni-O bond orderings. One is then left with a metallic state. Although weakened by electron doping, one may thus question if the *"electron-phonon coupling remains sufficiently large in the metallic regime for mediating the formation of Cooper pairs and producing the dome of T$_c$ versus doping content?"* We inspect this possibility in details in the following sections.

***Superconducting quantities:*** The electron-phonon coupling (EPC) is proportional to the inverse of the frequency square of modes, to the Reduced Electron-Phonon Coupling Matrix Element (REPME, labelled *D*) squared and to the density of states at the Fermi level N(E$_F$) (*i.e.* $\lambda \propto N(E_F)\frac{D^2}{\omega^2}$, see method). **Figure 8.a** displays the evolution of N(E$_F$) as a function of x extracted using a very dense k-mesh. The DOS at E$_F$ diminishes upon hole doping LaNiO$_2$ until it drops at x=0.25 (3d$^{8.75}$). The behavior between x=0 (3d$^9$) to x=0.1875 (3d$^{8.8125}$) agrees with expectation from a free electron system for which DOS evolves as $\sqrt{E}$.

We extract the Reduced Electron Phonon Coupling Matrix Element (REPME) induced by the bond disproportionation mode B$_{oc}$. To that end, we add an arbitrary amplitude Q$_{Boc}$ of 0.1106 Å/NiO$_2$ motif of the B$_{oc}$ mode – *i.e.* a displacement of 0.0553 Å per O atom in the ground state DSS structure of La$_{1-x}$Sr$_x$NiO$_2$ (x=0 to 0.25). We identify in the band structure a gap opening ΔE$_k$ of 0.64 eV along the Z-A path for x=0.1875 (**Appendix Figure 12**). The B$_{oc}$ mode induces a reduced electron-phonon matrix element (REPME) D=ΔE$_k$/Q$_{Boc}$=5.12 eV.Å$^{-1}$ (**see methods**). We checked that ΔE$_k$ scales linearly with u, *i.e.* D is constant for these amplitudes of vibration (**Appendix Figure 13**). We report in **Figure 8.b** the REPME as a function for the doping content. It describes a bell shape evolution whose maximal value is at x=0.1875 (3d$^{8.8125}$). This hints at the maximal T$_c$ observed experimentally for such doping content (x=0.2, 3d$^{8.8}$). Using the NM approximation as in Ref[15], we end up with a totally different behavior with a monotonously increasing D value with increasing x (see **Appendix Figure 9**). It nevertheless shows that extracting quantities within pristine compounds is not appropriated as doping acts on the REPME value. However, our SCAN-DFT simulations give a lower value for D. We have fitted the evolution of the D parameter between heterostructures at x=0.25, 0.125 and 0.0 in order to extract the improvement on $\Delta E_k$ produced by an HSE06 calculation over a SCAN calculation (**appendix Figure 14**). We observe a global improvement of 40% of the SCAN value. We thus rescale our $\Delta E_k$ on the PM LSNO by 1.4 in order to better describe this gap property (**Figure 8.b**).

Using the N($E_F$), $\omega_{Boc}^2$ and REPMEs with the HSE06 correction, we can compute the electron-phonon coupling for each doping content induced by the sole breathing mode (**Figure 8.c**). As one can see, $\lambda$ is dominated by the $D^2$ factor by describing a bell shape evolution with a coupling constant reaching 0.51 at x=0.1875 (3d$^{8.8125}$) of Sr doping content – in the charge ordered insulating phases, DOS at $E_F$ is supposedly zero and hence, there is no superconductivity. The value of $\lambda = 0.51$ is very close to the value extracted by the experimental analysis of Talantsev that identified $\lambda$ between 0.58 and 0.60 in the optimally doped nickelates ($R_{0.8}Sr_{0.2}NiO_2$, 3d$^{8.8}$)[76].

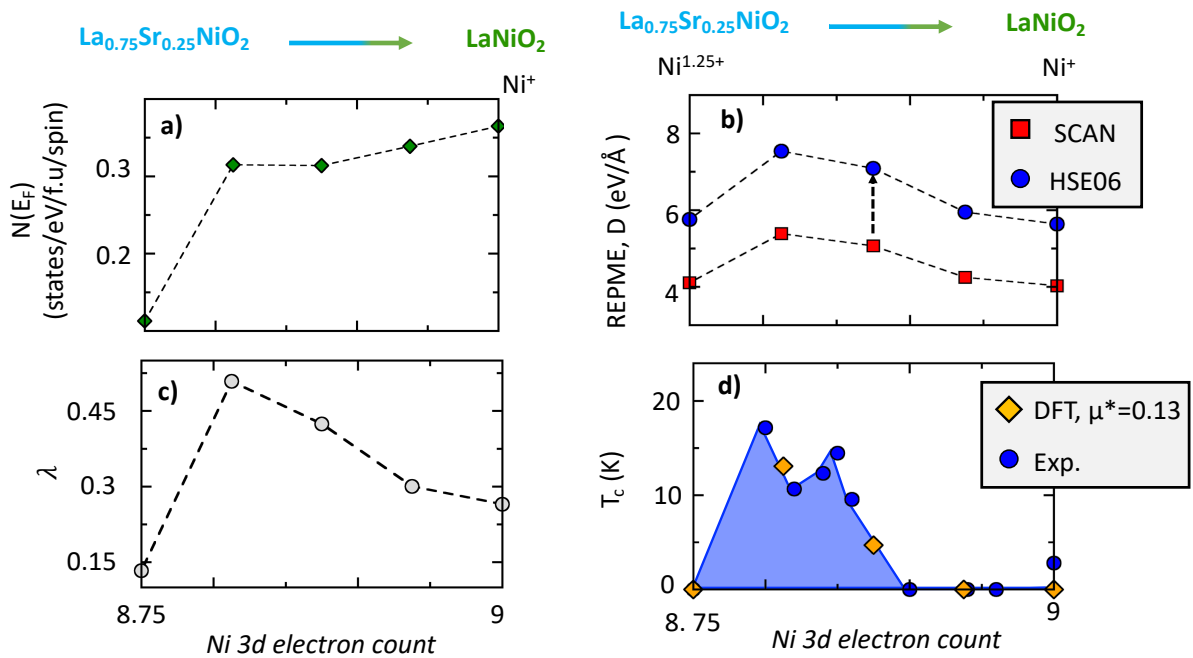

**Figure 8: Superconducting properties of doped nickelates.** *a) Density of states N($E_F$) (states/eV/f.u/spin channel) of the DSS nickelates upon Sr doping. b) Reduced Electron-Phonon Matrix Element (REPME, in eV.Å$^{-1}$) as a function of the Sr doping content obtained with SCAN (red filled squares) and rescaled by the HSE06 functional (blue filled circles). c) Computed electron-phonon coupling constant (no units) obtained from REPME, N($E_F$) and $\omega_{Boc}$. d) Computed critical temperature $T_c$ (in K) from our DFT simulations using usual screened Coulomb potential $\mu^*=0.13$ (orange filled diamonds) and $\mu=0.15$ (orange filled down triangles). Experimental values (blue filled circles) are*

*extracted from Ref.[6]. *1: The value at x=0 presents some uncertainty since the measurement of Ref. [6] was limited to 2 K and the resistivity did not reach 0.*

Using the Mac Millian equation[31] with the screened Coulomb potential µ* of 0.13 estimated from experimental data analysis of Talantsev[76] and assuming that $\omega_{Boc}$ is the characteristic energy scale for the interacting phonons contributing to the SC mechanism, our model yields a $T_c$ evolution as a function of Sr doping content in sharp agreement with the experimental values available in literature[6] (**Figure 8.d**). Numbers may of course be improved by involving higher level DFT functionals such as hybrid DFT or a full electron-phonon calculation implying all phonon modes – this is however not affordable for a 32 f.u with spin and cation disorders, no electron-phonon calculation technics allowing spin polarization is available up to date, to the best of our knowledge. However, the $T_c$ trend is already well captured by the SCAN functional, thereby suggesting that the bond disproportionation phonon mode is the dominating factor behind the superconducting properties of nickelates.

***The phonon driven mechanism is not incompatible with the experimental gap symmetry***: the superconducting gap symmetry is found to be non-trivial experimentally[77–79], pointing to an anisotropic gap with several pairing symmetries. This deviation from the most simple BCS isotropic *s*-wave pairing symmetry is usually understood to be a signature of non-conventional pairing. However, phonon mediated superconductivity does not impose any pairing symmetry: a phonon mediated mechanism and an anisotropic pairing can exist since the electron phonon coupling is momentum dependent. This has been highlighted in nickelates[80] as well as in cuprates[81]. Notably the pairing symmetry can be even nodal within a phonon mediated scheme and the evaluation of the critical temperature $T_c$ can be done with the same BCS type equation without loss of generality[82]. Thus, from the point of view of calculating the critical temperature, one may be able to evaluate with a good degree of accuracy using the usual *s*-wave equations for superconductors with any paring symmetry. From the point of view of the pairing symmetry, we can mainly refer to early theoretical works pointing to an s-wave paring symmetry can explain properly the experimental data of the London penetration depth and the superfluid current density[73]. This has been also highlighted in some of the experimental works[77] and in other global studies for several superconducting samples[76]. We can conclude here that our proposed model does not contradict the experimental data regarding the pairing symmetry and although we do not propose a complete picture of it, a *s*-wave phonon mediated scheme seems to be plausible and the easiest explanation.

**DISCUSSION**

The physical behavior and key quantities extracted with our DFT simulations in nickelates are reminiscent of trends in doping and superconducting properties of bismuth oxides superconductors [25,83]. Most notably, in both families, doping is a simple knob to progressively suppress an instability toward disproportionation effects producing an insulating state and to drive the material to the vicinity of a charge ordered state. Moreover, both compounds are single band system superconductors. Our calculations show that nickelates are phonon-driven superconductors and we thus conclude that they bear more similarities with bismuth and antimony oxide superconductors than with cuprates. We emphasize that a recent theoretical work also identifies an electron-phonon coupling mechanism albeit originating from distinct phonon modes[27]. This discrepancy could originate from the non-spin polarized (nonmagnetic, NM) approximation used in Ref.[27] . As we show here, the NM approximation (i) fails to describe phonons strongly coupled to local spin formations and (ii) produces a two bands system instead of a single band system within a real PM treatment. Consequently, a full electron-phonon calculation with local spin formation might be the desired approach to clarify these details. This is however not affordable since it would require full spin polarized calculation of the electron-phonon interaction which is not, to the best of our knowledge, implemented in any DFT code.

From our results, it appears that there are three important ingredients to reach superconductivity in nickelates: (i) getting a phonon that has to be soft enough so that it can get locally excited to produce attraction between the electrons, (ii) it has to produce a sufficiently large electron-phonon coupling constant and (iii) one needs a sufficiently large density of state at $E_F$ to get a large EPC. Since DFT that only treats static mean-field correlation effects is a sufficient platform to get relevant quantities and information on the superconducting mechanism, it suggests that dynamical correlation effects as codified by the Hubbard model play a marginal role in the physics of nickelates. However, we do not claim that any other forms of correlations are not important since allowing local spin formation is an essential factor to identify the breathing distortion.

Finally, undoped materials are not necessarily a good starting point for understanding trends in doping effects and superconducting properties in oxides. The search of potential electronic and structural instabilities in the doped phase diagram appears as a critical aspect. Once the proper starting point is established, the identified phonon driven mechanism common between nickel oxide and bismuth oxide superconductors therefore calls for new investigations of doping effects from the point of view of structural distortions and intrinsic electronic instabilities rather than from the strict point of view of electron correlation in other oxide superconductors.

**Acknowledgements**

This work has received financial support from the CNRS through the MITI interdisciplinary programs under the project SuNi and through the ANR SUPERNICKEL. Authors acknowledge access granted to HPC resources of Criann through the projects 2020005 and 2007013 and of Cines through the DARI project A0080911453. L. I. acknowledges the funding from the Ile de France region and the European Union's Horizon 2020 research and innovation programme under the Marie Sklodowska-Curie grant agreement №21004513 (DOPNICKS project).

*References*

[1] L.N. Cooper, Phys. Rev. **104**, 1189 (1956).

[2] J.G. Bednorz and K.A. Müller, Z. Phys. B - Condens. Matter **64**, 189 (1986).

[3] T.M. Rice, Phys. Rev. B **59**, 7901 (1999).

[4] J. Chaloupka and G. Khaliullin, Phys. Rev. Lett. **100**, 16404 (2008).

[5] D. Li, K. Lee, B.Y. Wang, M. Osada, S. Crossley, H.R. Lee, Y. Cui, Y. Hikita, and H.Y. Hwang, Nature **572**, 624 (2019).

[6] M. Osada, B.Y. Wang, B.H. Goodge, S.P. Harvey, K. Lee, D. Li, L.F. Kourkoutis, and H.Y. Hwang, Adv. Mater. **33**, 2104083 (2021).

[7] M. Osada, B.Y. Wang, B.H. Goodge, K. Lee, H. Yoon, K. Sakuma, D. Li, M. Miura, L.F. Kourkoutis, H.Y. Hwang, M. Osada, and H.Y. Hwang, Nano Lett. **20**, 5735 (2020).

[8] S. Zeng, C. Li, L.E. Chow, Y. Cao, Z. Zhang, and C.S. Tang, Sci. Adv. **8**, eabl9927 (2022).

[9] G.A. Pan, D.F. Segedin, H. LaBollita, Q. Song, E.M. Nica, B.H. Goodge, A.T. Pierce, S. Doyle, S. Novakov, D.C. Carrizales, A.T. N'Diaye, P. Shafer, H. Paik, J.T. Heron, J.A. Mason, A. Yacoby, L.F. Kourkoutis, O. Erten, C.M. Brooks, A.S. Botana, and J.A. Mundy, Nat. Mater. **21**, 160 (2021).

[10] Y. Maeno, H. Hashimoto, K. Yoshida, S. Nishizaki, T. Fujita, J.G. Bednorz, and F. Lichtenberg, Nature **372**, 532 (1994).

[11] Y.J. Yan, M.Q. Ren, H.C. Xu, B.P. Xie, R. Tao, H.Y. Choi, N. Lee, Y.J. Choi, T. Zhang, and D.L. Feng, Phys. Rev. X **5**, 041018 (2015).

[12] Y.K. Kim, N.H. Sung, J.D. Denlinger, and B.J. Kim, Nat. Phys. **12**, 37 (2016).

[13] X. Zhou, X. Zhang, J. Yi, P. Qin, Z. Feng, P. Jiang, Z. Zhong, H. Yan, X. Wang, H. Chen, H. Wu, X. Zhang, Z. Meng, X. Yu, M.B.H. Breese, J. Cao, J. Wang, C. Jiang, and Z. Liu, Adv. Mater. **34**, 2106117 (2022).

[14] F. Lechermann, Phys. Rev. B **101**, 08110 (2020).

[15] Y. Nomura, M. Hirayama, T. Tadano, Y. Yoshimoto, K. Nakamura, and R. Arita, Phys. Rev. B **100**, 205138 (2019).

[16] M. Hirayama, T. Tadano, Y. Nomura, and R. Arita, Phys. Rev. B **101**, 075107 (2020).

[17] V. Christiansson, F. Petocchi, and P. Werner, Phys. Rev. B **107**, 045144 (2023).

[18] T.Y. Xie, Z. Liu, C. Cao, Z.F. Wang, J.L. Yang, and W. Zhu, Phys. Rev. B **106**, 035111 (2022).

[19] Z. Liu, Z. Ren, W. Zhu, Z. Wang, and J. Yang, Npj Quantum Mater. **5**, 31 (2020).

[20] P. Adhikary, S. Bandyopadhyay, T. Das, I. Dasgupta, and T. Saha-Dasgupta, Phys. Rev. B **102**, 100501(R) (2020).

[21] F. Bernardini, V. Olevano, and A. Cano, Phys. Rev. Res. **2**, 013219 (2020).

[22] M. Hepting, D. Li, C.J. Jia, H. Lu, E. Paris, Y. Tseng, X. Feng, M. Osada, E. Been, Y. Hikita, Y.D. Chuang, Z. Hussain, K.J. Zhou, A. Nag, M. Garcia-Fernandez, M. Rossi, H.Y. Huang, D.J. Huang, Z.X. Shen, T. Schmitt, H.Y. Hwang, B. Moritz, J. Zaanen, T.P. Devereaux, and W.S. Lee, Nat. Mater. **19**, 381 (2020).

[23] A.W. Sleight, Phys. C Supercond. Its Appl. **514**, 152 (2015).

[24] C.H.P. Wen, H.C. Xu, Q. Yao, R. Peng, X.H. Niu, Q.Y. Chen, Z.T. Liu, D.W. Shen, Q. Song, X. Lou, Y.F. Fang, X.S. Liu, Y.H. Song, Y.J. Jiao, T.F. Duan, H.H. Wen, P. Dudin, G. Kotliar, Z.P. Yin, and D.L. Feng, Phys. Rev. Lett. **121**, 117002 (2018).

[25] J. Varignon, Npj Comput. Mater. **9**, 30 (2023).

[26] M. Kim, G.M. McNally, H.H. Kim, M. Oudah, A.S. Gibbs, P. Manuel, R.J. Green, R. Sutarto, T. Takayama, A. Yaresko, U. Wedig, M. Isobe, R.K. Kremer, D.A. Bonn, B. Keimer, and H. Takagi, Nat. Mater. **21**, 627 (2022).

[27] Z. Li and S.G. Louie, ArXiv **2210.12819**, (2022).

[28] J. Varignon, M. Bibes, and A. Zunger, Nat. Commun. **10**, 1658 (2019).

[29] Z.P. Yin, A. Kutepov, and G. Kotliar, Phys. Rev. X **3**, 021011 (2013).

[30] J.M. An and W.E. Pickett, Phys. Rev. Lett. **86**, 4366 (2001).

[31] P.B. Allen and R. C. Dynes, Phys. Rev. B **12**, 905 (1975).

[32] J. Varignon, M. Bibes, and A. Zunger, Phys. Rev. B **100**, 035119 (2019).

[33] J. Sun, A. Ruzsinszky, and J. Perdew, Phys. Rev. Lett. **115**, 036402 (2015).

[34] L. Iglesias, M. Bibes, and J. Varignon, Phys. Rev. B **104**, 035123 (2021).

[35] A.A. Carrasco Alvarez, M. Bibes, W. Prellier, and J. Varignon, Phys. Rev. B **107**, 115109 (2023).

[36] A. V. Krukau, O.A. Vydrov, A.F. Izmaylov, and G.E. Scuseria, J. Chem. Phys. **125**, 224106 (2006).

[37] K. Pokharel, C. Lane, J.W. Furness, R. Zhang, J. Ning, B. Barbiellini, R.S. Markiewicz, Y. Zhang, A. Bansil, and J. Sun, Npj Comput. Mater. **8**, 31 (2022).

[38] J.K. Perry, J. Tahir-Kheli, and W.A. Goddard, Phys. Rev. B **65**, 1445011 (2002).

[39] A. Van de Walle, M. Asta, and G. Ceder, Calphad Comput. Coupling Phase Diagrams Thermochem. **26**, 539 (2002).

[40] A. Zunger, S.-H. Wei, L.G. Ferreira, and J.E. Bernard, Phys. Rev. Lett. **65**, 353 (1990).

[41] B.J. Campbell, H.T. Stokes, D.E. Tanner, and D.M. Hatch, J. Appl. Crystallogr. **39**, 607 (2006).

[42] ISOTROPY Software Suite, iso.byu.edu. Available at:

https://iso.byu.edu/iso/isodistort_version5.6.1/isodistort.php.

[43] J. Varignon, O.I. Malyi, and A. Zunger, Phys. Rev. B **105**, 165111 (2022).

[44] G. Kresse and J. Haffner, Phys. Rev. B **47**, 558 (1993).

[45] G. Kresse and J. Furthmüller, Comput. Mater. Sci. **6**, 15 (1996).

[46] P.E. Blöchl, Phys. Rev. B **50**, 17953 (1994).

[47] A.A. Mostofi, J.R. Yates, Y.-S. Lee, I. Souza, D. Vanderbilt, and N. Marzari, Comput. Phys. Commun. **178**, 685 (2008).

[48] N. Marzari and D. Vanderbilt, Phys. Rev. B **56**, 12847 (1997).

[49] I. Souza, N. Marzari, and D. Vanderbilt, Phys. Rev. B **65**, 35109 (2001).

[50] G. Pizzi, V. Vitale, R. Arita, S. Blügel, F. Freimuth, G. Géranton, M. Gibertini, D. Gresch, C. Johnson, T. Koretsune, J. Ibañez-Azpiroz, H. Lee, J.M. Lihm, D. Marchand, A. Marrazzo, Y. Mokrousov, J.I. Mustafa, Y. Nohara, Y. Nomura, L. Paulatto, S. Poncé, T. Ponweiser, J. Qiao, F. Thöle, S.S. Tsirkin, M. Wierzbowska, N. Marzari, D. Vanderbilt, I. Souza, A.A. Mostofi, and J.R. Yates, J. Phys. Condens. Matter **32**, 165902 (2020).

[51] Z. Chen, M. Osada, D. Li, E.M. Been, S. Di Chen, M. Hashimoto, D. Lu, S.K. Mo, K. Lee, B.Y. Wang, F. Rodolakis, J.L. McChesney, C. Jia, B. Moritz, T.P. Devereaux, H.Y. Hwang, and Z.X. Shen, Matter **5**, 1806 (2022).

[52] B.H. Goodge, D. Li, K. Lee, M. Osada, B.Y. Wang, G.A. Sawatzky, H.Y. Hwang, and L.F. Kourkoutis, Proc. Natl. Acad. Sci. U. S. A. **118**, e2007683118 2020 (2021).

[53] M.Y. Choi, K.W. Lee, and W.E. Pickett, Phys. Rev. B **101**, 20503 (2020).

[54] R. Zhang, C. Lane, B. Singh, J. Nokelainen, B. Barbiellini, R.S. Markiewicz, A. Bansil, and J. Sun, Commun. Phys. **4**, 118 (2021).

[55] M.Y. Choi, W.E. Pickett, and K.W. Lee, Phys. Rev. Res. **2**, 033445 (2020).

[56] J. Varignon, M.N. Grisolia, J. Íñiguez, A. Barthélémy, and M. Bibes, Npj Quantum Mater. **2**, 21 (2017).

[57] A. Mercy, J. Bieder, J. Íñiguez, and P. Ghosez, Nat. Commun. **8**, 1 (2017).

[58] G.M. Dalpian, Q. Liu, J. Varignon, M. Bibes, and A. Zunger, Phys. Rev. B **98**, 075135 (2018).

[59] V. V. Poltavets, K.A. Lokshin, S. Dikmen, M. Croft, T. Egami, and M. Greenblatt, J. Am. Chem. Soc. **128**, 9050 (2006).

[60] K.A. Lokshin, D. Mitchell, M. V. Lobanov, V. Struzhkin, and T. Egami, ECS J. Solid State Sci. Technol. **11**, 044008 (2022).

[61] V. Pardo and W.E. Pickett, Phys. Rev. B - Condens. Matter Mater. Phys. **83**, 245128 (2011).

[62] Y. Shen, J. Sears, G. Fabbris, J. Li, J. Pelliciari, I. Jarrige, X. He, I. Božović, M. Mitrano, J. Zhang, J.F. Mitchell, A.S. Botana, V. Bisogni, M.R. Norman, S. Johnston, and M.P.M. Dean, Phys. Rev. X **12**, 1 (2022).

[63] J. Zhang, Y.S. Chen, D. Phelan, H. Zheng, M.R. Norman, and J.F. Mitchell, Proc. Natl. Acad. Sci. U. S. A. **113**, 8945 (2016).

[64] X. Chen, H. Zheng, D.P. Phelan, H. Zheng, S.H. Lapidus, M.J. Krogstad, R. Osborn, S. Rosenkranz, and J.F. Mitchell, Chem. Mater. **34**, 4560 (2022).

[65] V. V. Poltavets, K.A. Lokshin, A.H. Nevidomskyy, M. Croft, T.A. Tyson, J. Hadermann, G. Van Tendeloo, T. Egami, G. Kotliar, N. Aproberts-Warren, A.P. Dioguardi, N.J. Curro, and M. Greenblatt, Phys. Rev. Lett. **104**, 6 (2010).

[66] A.S. Botana, V. Pardo, W.E. Pickett, and M.R. Norman, Phys. Rev. B **94**, 081105(R) (2016).

[67] K.G. Slobodchikov and I. V. Leonov, Phys. Rev. B **106**, 165110 (2022).

[68] G. Krieger, L. Martinelli, S. Zeng, L.E. Chow, K. Kummer, R. Arpaia, M. Moretti Sala, N.B. Brookes, A. Ariando, N. Viart, M. Salluzzo, G. Ghiringhelli, and D. Preziosi, Phys. Rev. Lett. **129**, 27002 (2022).

[69] A. Raji, G. Krieger, N. Viart, D. Preziosi, J.-P. Rueff, and A. Gloter, Small 2304872 (2023).

[70] A. Mercy, J. Bieder, J. Íñiguez, and P. Ghosez, Nat. Commun. **8**, 1677 (2017).

[71] H. Chen, A. Hampel, J. Karp, F. Lechermann, and A.J. Millis, Front. Phys. **10**, 835942 (2022).

[72] R. Jiang, Z.-J. Lang, T. Berlijn, and W. Ku, ArXiv **2207.08802**, (2022).

[73] E.F. Talantsev, Results Phys. **17**, 103118 (2020).

[74] C. Lane, R. Zhang, B. Barbiellini, R.S. Markiewicz, A. Bansil, J. Sun, and J.X. Zhu, Commun. Phys. **6**, 90 (2023).

[75] R. Zhang, C. Lane, J. Nokelainen, B. Singh, B. Barbiellini, R.S. Markiewicz, A. Bansil, and J. Sun, ArXiv 2207.00784 (2022).

[76] E.F. Talantsev, ArXiv 302.14729 (2023).

[77] L.E. Chow, S.K. Sudheesh, P. Nandi, S.W. Zeng, Z.T. Zhang, X.M. Du, Z.S. Lim, E.E.M. Chia, and A. Ariando, ArXiv 2201.10038 (2022).

[78] Q. Gu, Y. Li, S. Wan, H. Li, W. Guo, H. Yang, Q. Li, X. Zhu, X. Pan, Y. Nie, and H.H. Wen, Nat. Commun. **11**, 6027 (2020).

[79] S.P. Harvey, B.Y. Wang, J. Fowlie, M. Osada, K. Lee, Y. Lee, D. Li, and H.Y. Hwang, ArXiv: 2201.12971 (2022).

[80] X. Sui, J. Wang, X. Ding, K.-J. Zhou, L. Qiao, H. Lin, and B. Huang, ArXiv 2202.11904 (2022).

[81] T.P. Devereaux, T. Cuk, Z.X. Shen, and N. Nagaosa, Phys. Rev. Lett. **93**, 117004 (2004).

[82] P.W. Anderson and P. Morel, Phys. Rev. **123**, 1911 (1961).

[83] A.W. Sleight, Phys. C Supercond. Its Appl. **514**, 152 (2015).

**APPENDICES**

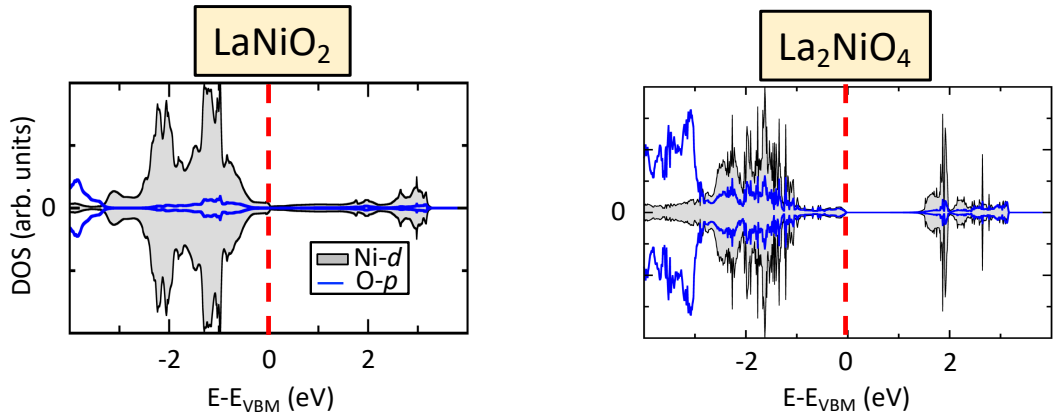

***APPENDIX FIGURE 1***: *density of states of LaNiO$_2$ and La$_2$NiO$_4$ projected on the Ni d (grey area) and O p (blue line) states. The paramagnetic order Is used.*

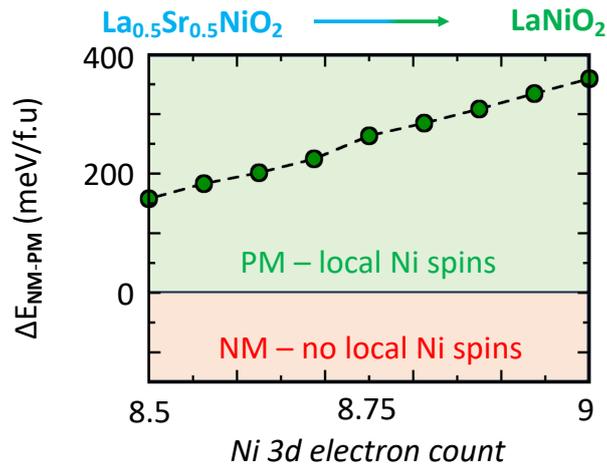

***APPENDIX FIGURE 2:*** *Energy difference (in meV/f.u) between the non-spin polarized (NM) and the paramagnetic (PM) ground state as a function of the formal Ni valency.*

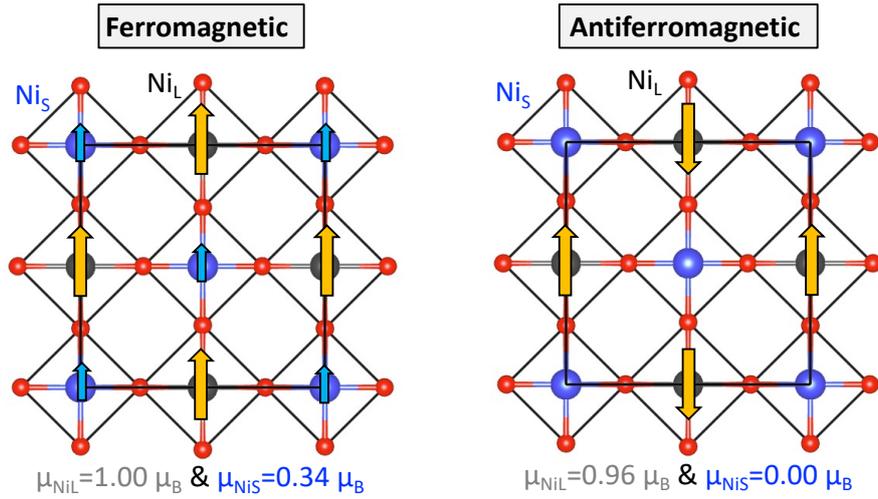

$\mu_{NiL}=1.00\ \mu_B$ & $\mu_{NiS}=0.34\ \mu_B$ $\quad\quad$ $\mu_{NiL}=0.96\ \mu_B$ & $\mu_{NiS}=0.00\ \mu_B$

**APPENDIX FIGURE 3:** Effect of the local potential experienced by $Ni_S$ cations. Magnetic moments computed on $Ni_S$ cations for a ferromagnetic configuration (**a**) or for an antiferromagnetic arrangement (**b**) of $Ni_L$ cation in the (xy) plane. The structure is set the disproportionated material identified at half doping for a $(LaNiO_2)_1/(SrNiO_2)_1$ heterostructure. The SCAN functional is used.

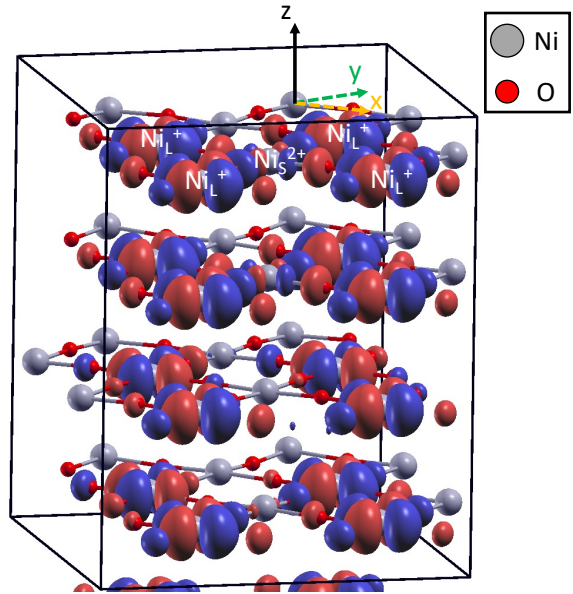

***APPENDIX FIGURE 4:*** *Wannier functions associated built within the $(LaNiO_2)_1/(SrNiO_2)_1$ superlattice approach and with a PM order. Only occupied levels are considered.*

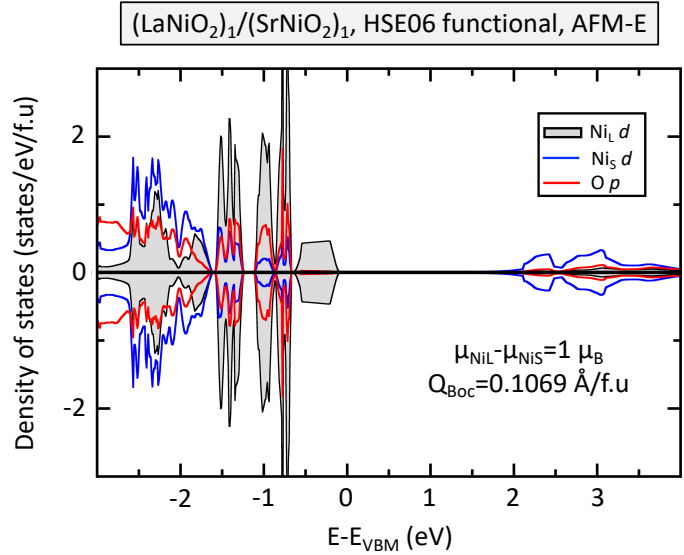

**APPENDIX FIGURE 5:** Density of states associates with a (LaNiO$_2$)$_1$/(SrNiO$_2$)$_1$ superlattice using the HSE06 hybrid functional. An AFME order is used for this simulation.

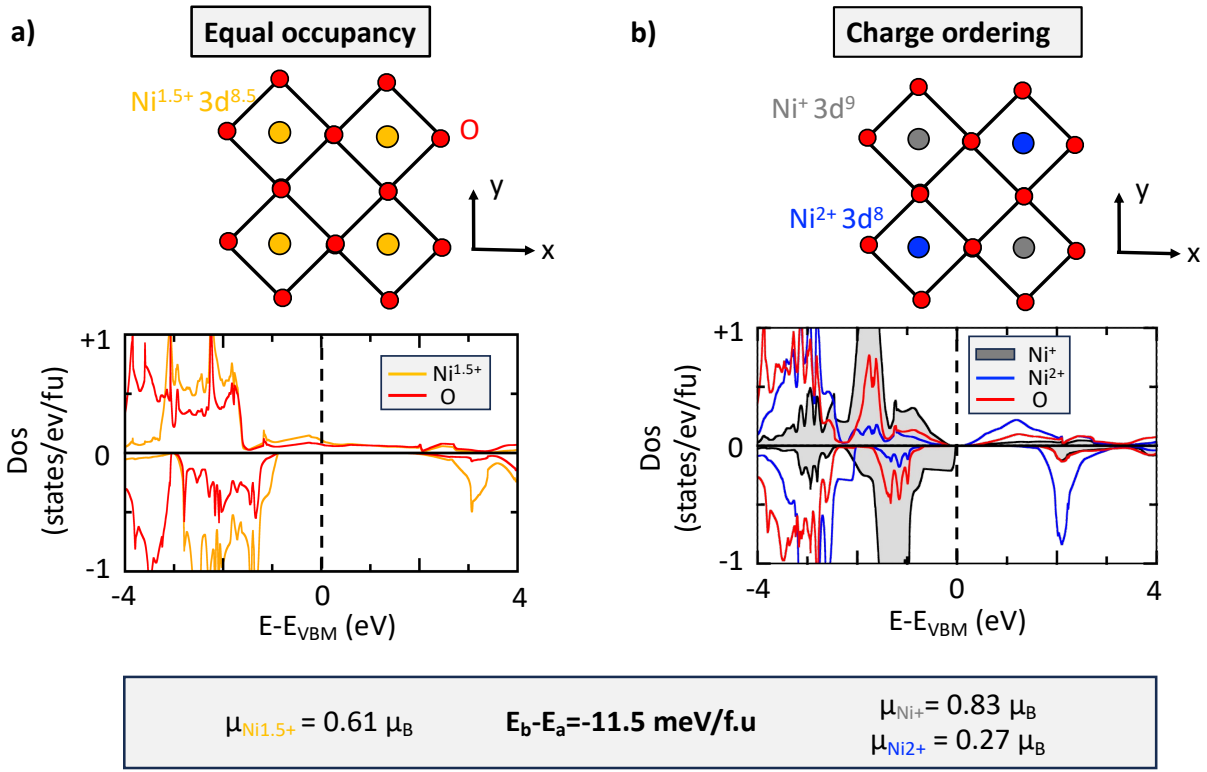

**APPENDIX FIGURE 6:** Electronic instability associated with the 1.5+ formal oxidation state of Ni cations. Total energy calculation of a La$_{0.5}$Sr$_{0.5}$NiO$_2$ plane with 2 formula units in the (xy) plane in which one enforces similar occupancy on Ni cations (a) and a 2D checkerboard pattern of Ni$^+$ anf Ni$^{2+}$ cations. The structure is fixed to undistorted P4/mmm cell, a ferromagnetic order is used and the HSE06 calculation is employed with a cut-off to 650 eV. The lower panels represent the projected density of states

(states/eV/cell/f.u) projected on Ni cations and O p states. The HSE06 functional is used for these results.

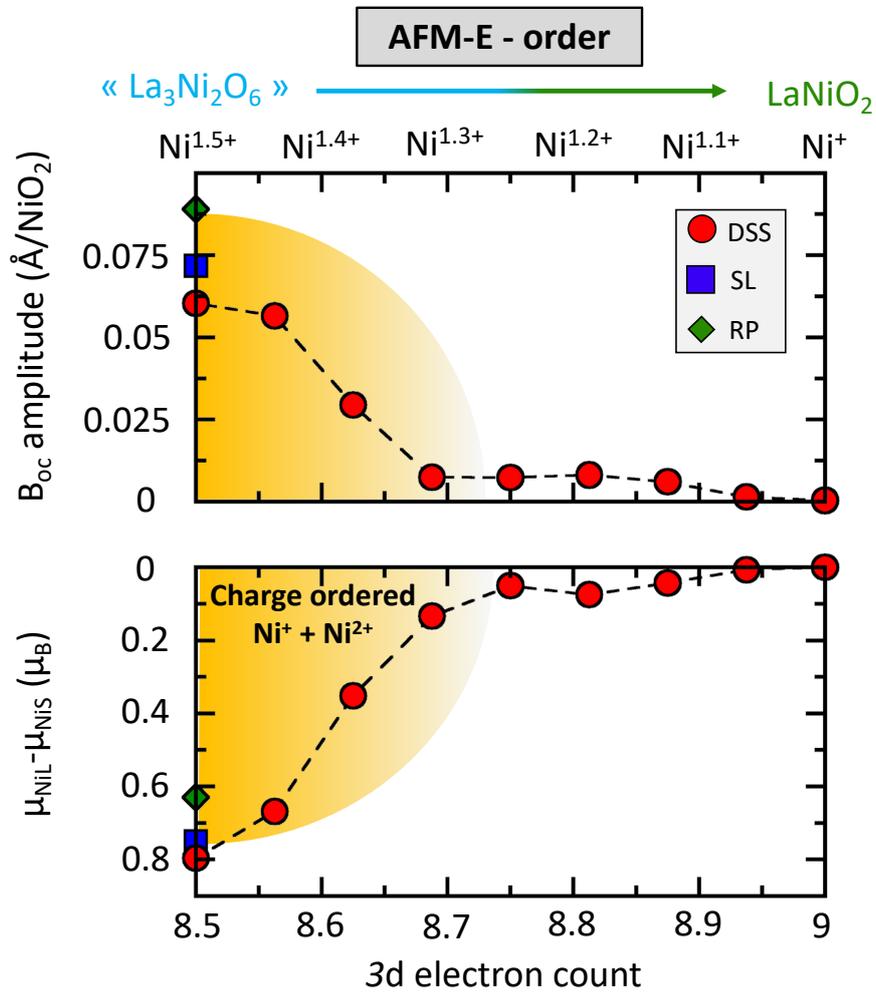

**APPENDIX FIGURE 7**: Bond disproportionation amplitude (in /fu) (upper panel) and magnetic moment asymmetry (in $\mu_B$) as a function of the doping content. Results are obtained by using an AFM-E order consisting of up-up-down-down spin chains in the NiO$_2$ planes with FM couplings along the c axis.

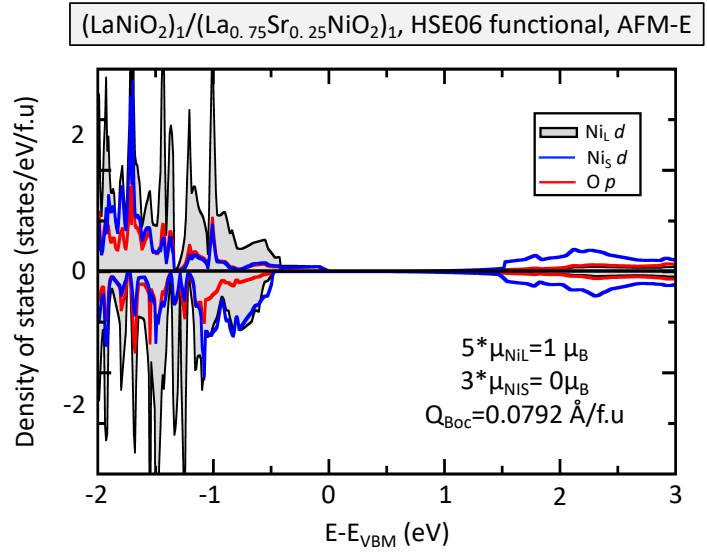

**APPENDIX FIGURE 8:** *Density of states associates with a (LaNiO2)1/(La$_{0.75}$Sr$_{0.25}$NiO2)1 superlattice using HSE06 hybrid functional. An AFM-E order is used through the calculation.*

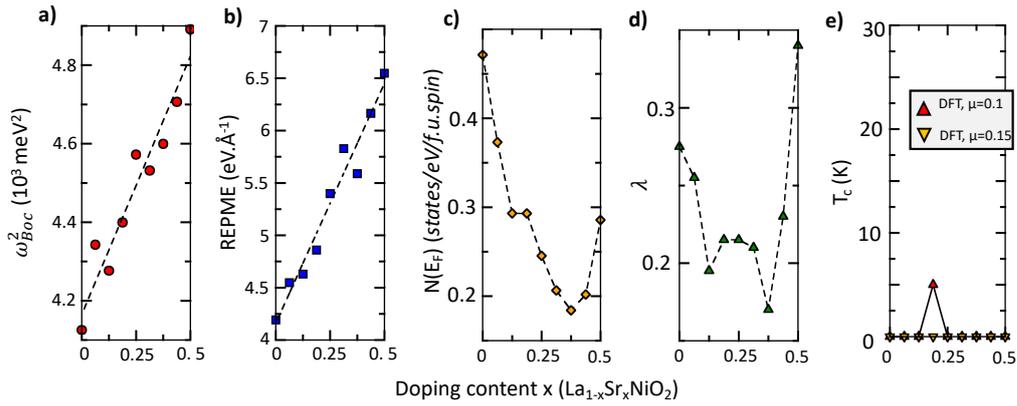

**APPENDIX FIGURE 9:** *Superconducting quantities using SCAN and NM solutions as a function of the doping content. **a)** Squared frequency (in meV$^2$) of the B$_{oc}$ mode starting from a P$_4$/mmm cell. **b)** Computed Reduced Electron Phonon matrix Element (REPME, in eV/Å). **c)** Density of states at E$_F$ N(E$_F$) at the Femi level E$_F$ associated with the single band affected by the B$_{oc}$ mode. **d)** Electron phonon coupling constant deduced from N(E$_F$), $\omega^2$ and the REPMEs. **e)** Computed critical temperature for usual screened Coulomb potential µ of 0.1 (red filled up triangles) and 0.15 (orange filled down triangles).*

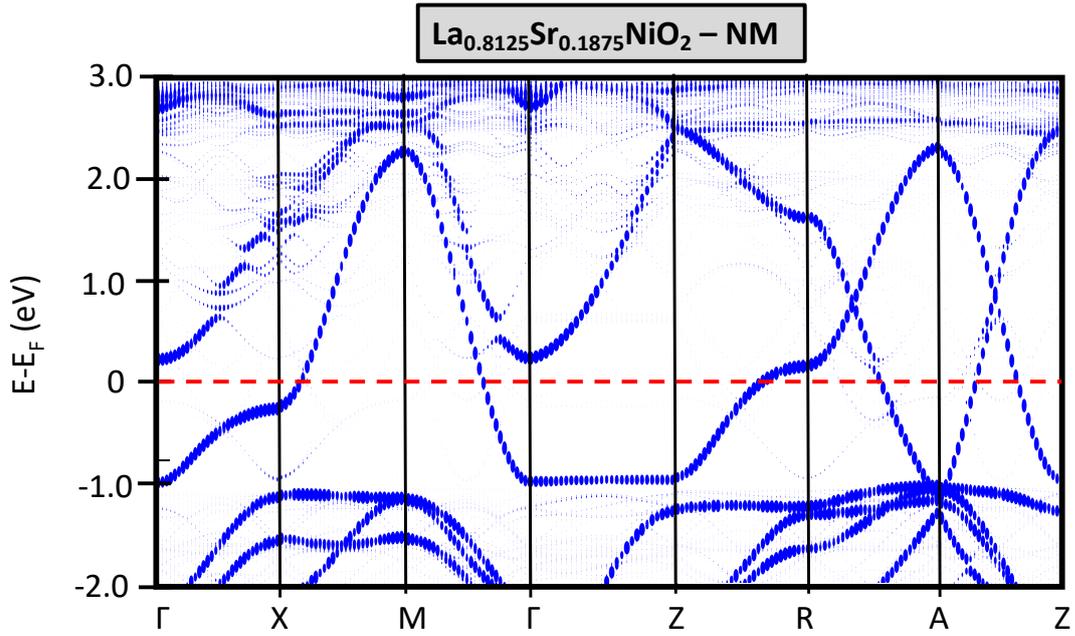

***APPENDIX FIGURE 10:*** *Unfolded band structure of La$_{0.8125}$Sr$_{0.1875}$NiO$_2$ DSS using the SCAN functional and a NM solution.*

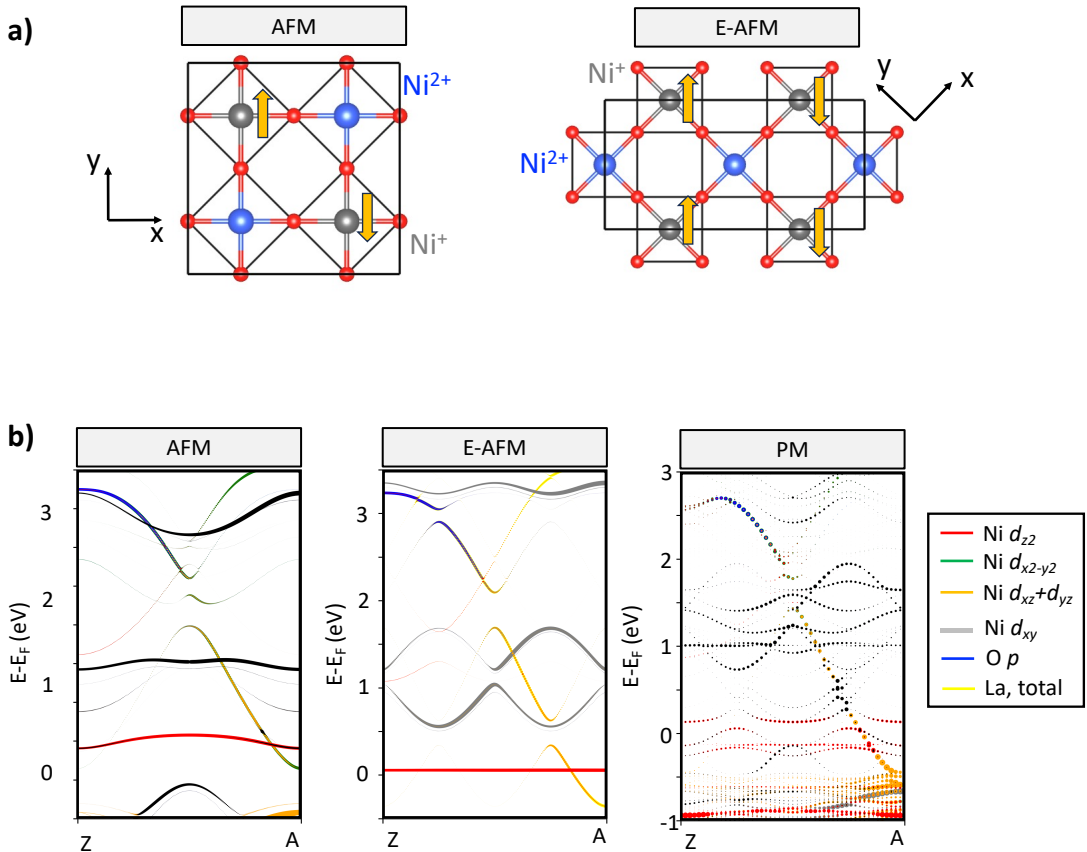

***APPENDIX FIGURE 11:*** *Band dispersion of a 50% doped nickelate superlattice for various magnetic orderings. **a)** Sketches of long range antiferromagnetic orders used for the band dispersions. **b)***

*Unfolded band dispersion to the primitive $P_4/mmm$ Brillouin zone for the antiferromagnetic (AFM), E-type antiferromagnetic (E-AFM) and paramagnetic (PM) orders along the Z-A path. The coordinates of the points are Z (0,0,1/2) and A (1/2, 1/2 , 1/2). The SCAN functional is used. Bands are projected on atomic and orbital characters.*

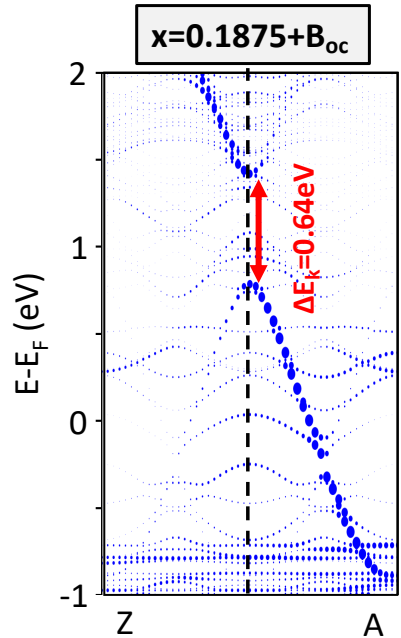

**APPENDIX FIGURE 12:** *Unfolded band structure when adding an amplitude $Q_{Boc}$ of 0.1106 Å/$NiO_2$ motif in the $La_{0.8125}Sr_{0.1875}NiO_2$ DSS using the SCAN functional and a PM order. The coordinates of the Brillouin zone points are Z (0,0,1/2) and A (1/2,1/2,1/2).*

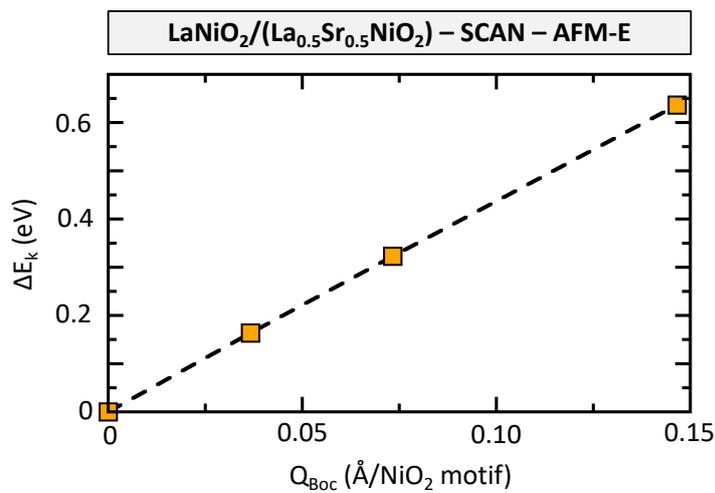

**APPENDIX FIGURE 13:** *Evolution of the gap opening ΔE$_k$ forced by a given amplitude of the B$_{oc}$ mode in the band structure for a model La$_{0.75}$Sr$_{0.25}$NiO$_2$ system with an AFM-E order. Results are obtained with the SCAN functional.*

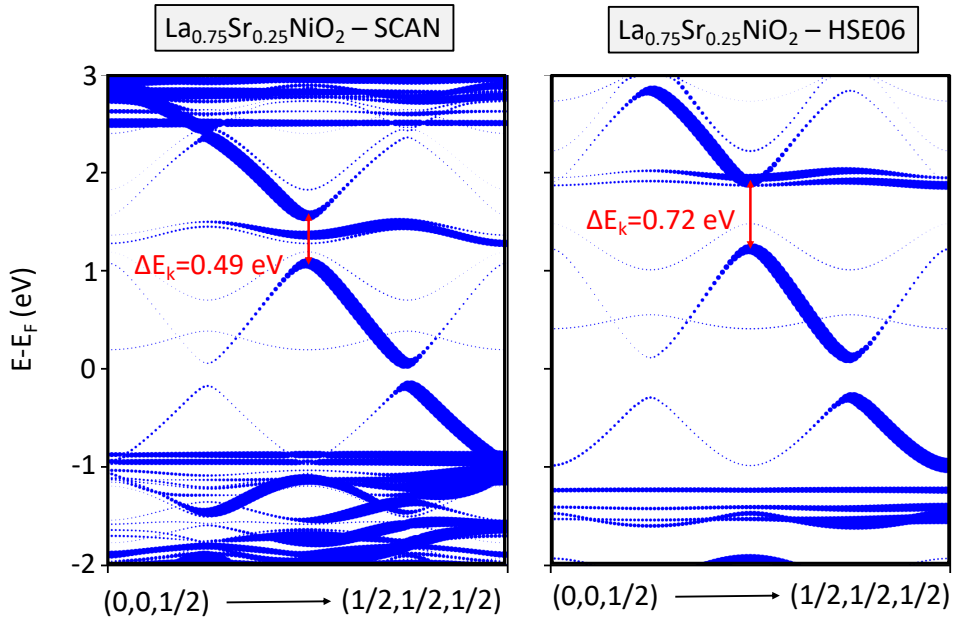

**APPENDIX FIGURE 14:** *Gap opening ΔE$_k$ induced by a fixed amplitude of the B$_{oc}$ bond disproportionation mode frozen in the structure using a semi-local SCAN functional (left panel) and a more evolved hybrid HSE06 functional (right panel). The improvement of HSE06 over SCAN gap and hence REPME is estimated to 47%. A model heterostructure of the form LaNiO$_2$/(La$_{0.5}$Sr$_{0.5}$NiO$_2$) with an AFM-C order (AFM coupling between nearest neighbor Ni cations) is used for these simulations. The very same calculation but using a FM order yields an improvement of 31% for ΔE$_k$ of the HSE06 over the SCAN functional. Assuming that within a PM, we would have 50% of FM and 50% of AFM interactions in plane, we apply a correction of 39% to our SCAN-PM gap openings.*